\documentclass[letter]{aa}  

\usepackage{txfonts}
\usepackage{hyperref,longtable}

\usepackage{silence}
\usepackage{graphicx}
\graphicspath{{./Fig_}}
\usepackage{transparent}
\usepackage{xkeyval,xcolor}
\makeatletter
\newlength{\sfp@hseplen}\newlength{\sfp@vseplen}
\define@cmdkey{subfigpos}[sfp@]{pos}[ul]{}
\define@cmdkey{subfigpos}[sfp@]{font}[\small]{}
\define@cmdkey{subfigpos}[sfp@]{vsep}[0.75\baselineskip]{\setlength{\sfp@vseplen}{\sfp@vsep}}
\define@cmdkey{subfigpos}[sfp@]{hsep}[3.5pt]{\setlength{\sfp@hseplen}{\sfp@hsep}}
\newcommand{\subfigimg}[4][,]{%
        \setkeys{Gin,subfigpos}{pos,font,vsep,hsep,#1}
        \setbox1=\hbox{\includegraphics{#4}}
        \ifnum\pdfstrcmp{\sfp@pos}{ul}=0
                \leavevmode\rlap{\usebox1}
                \rlap{\hspace*{\sfp@hsep}\raisebox{\dimexpr\ht1-\sfp@vsep}{\transparent{#3}{\setlength{\fboxsep}{1pt}\colorbox{white}{%
\transparent{1}\sfp@font{#2}}}%
}}
                \phantom{\usebox1}
        \else\ifnum\pdfstrcmp{\sfp@pos}{ur}=0
                \leavevmode\usebox1
                \llap{\raisebox{\dimexpr\ht1-\sfp@vsep}{\sfp@font{#2}}\hspace*{\sfp@hsep}}
        \else\ifnum\pdfstrcmp{\sfp@pos}{lr}=0
                \leavevmode\usebox1
                \llap{\raisebox{\sfp@vsep}{\sfp@font{#2}}\hspace*{\sfp@hsep}}
        \else
                \leavevmode\rlap{\usebox1}
                \rlap{\hspace*{\sfp@hseplen}\raisebox{\sfp@vsep}{\sfp@font{#2}}}
                \phantom{\usebox1}
        \fi\fi\fi
}
\newcommand{\fontfig}[1]{\tiny$\!\!$\color{#1}\textbf}
\newcommand{\AspectRatio}[1]{\dimexpr 1pt * \wd#1 / \ht#1 \relax} 

\usepackage{amsmath} 
\usepackage{array}
\usepackage{adjustbox}
\usepackage{multirow}
\usepackage{makecell}
\newcolumntype{C}[1]{>{\centering\arraybackslash}p{#1}} 

\usepackage{placeins}

\usepackage{algorithm}
\usepackage[noend]{algpseudocode}
\newcommand{\commentalgo}[1]{\Comment{{\tiny #1}}} 

\newcommand{\eq}[1]{Eq.~\eqref{#1}\xspace}
\newcommand{\eqs}[2]{Eqs.~(\ref{#1},\ref{#2})\xspace}

\newcommand{\subfigref}[1]{(#1)} 
\newcommand{\fig}[1]{Fig.~\ref{#1}\xspace}
\newcommand{\figfull}[1]{Figure~\ref{#1}\xspace} 
\newcommand{\subfig}[2]{Fig.~\ref{#1}\subfigref{#2}\xspace}
\newcommand{\subfigfull}[2]{Figure~\ref{#1}\subfigref{#2}\xspace} 
\newcommand{\subfigs}[2]{Figs.~\ref{#1}\subfigref{#2}\xspace}

\newcommand{\tab}[1]{Table~\ref{#1}\xspace}

\newcommand{\refsec}[1]{Section~\ref{#1}\xspace} 
\newcommand{\refapp}[1]{Appendix~\ref{#1}\xspace} 

\usepackage{mathtools}
\usepackage{mathrsfs}
\usepackage{nicefrac}

\newcommand{\Tag}[1]{\text{#1}}        

\newcommand{\V}[1]{{\boldsymbol{#1}}}                 

\newcommand{\E}[1]{\mathrm{e}^{#1}}                   

\DeclarePairedDelimiterX{\paren}[1]{(}{)}{#1}
\newcommand{\Paren}[1]{\paren*{#1}}
\let\brace=\undefined 
\DeclarePairedDelimiterX{\brace}[1]{\{}{\}}{#1}

\let\brack=\undefined 
\DeclarePairedDelimiterX{\brack}[1]{[}{]}{#1}
\newcommand{\Brack}[1]{\brack*{#1}}
\DeclarePairedDelimiterX{\abs}[1]{\rvert}{\lvert}{#1}     
\newcommand{\Abs}[1]{\abs*{#1}}
\DeclarePairedDelimiterX{\norm}[1]{\lVert}{\rVert}{#1}    
\newcommand{\Norm}[1]{\norm*{#1}}

\DeclarePairedDelimiterX{\avg}[1]{\langle}{\rangle}{#1}   
\newcommand{\Avg}[1]{\avg*{#1}}

\DeclarePairedDelimiterX{\ceil}[1]{\lceil}{\rceil}{#1}     

\DeclarePairedDelimiterX{\floor}[1]{\lfloor}{\rfloor}{#1}  

\newcommand{\conv}{\star}                                                               

\newcommand{\x}{{\V{\theta}}}                                   
\newcommand{\sig}{{\V{\sigma}}}                                 
\newcommand{\Vcube}{{\V{c}^{\Tag{red}}}}                
\newcommand{\Vdata}{{\V{d}}}                                    
\newcommand{\Vobj}{{\V{o}}}                                             
\newcommand{\Vpsf}{{\V{p}}}                                             

\newcommand{\Vred}{{\Vdata^{\Tag{red}}}}                
\newcommand{\Vbg}{{\Vdata^{\Tag{bg}}}}                  
\newcommand{\Vmoon}{{\Vdata^{\Tag{moon}}}}              
\newcommand{\Vdec}{{\Vobj^{\Tag{dec}}}}                 
\newcommand{\Vthr}{{\Vobj^{\Tag{thr}}}}                 
\newcommand{\Vhalo}{{\Vdata^{\Tag{hal}}}}               
\newcommand{\Vres}{{\Vdata^{\Tag{res}}}}                
\newcommand{\Vfilt}{{\Vdata^{\Tag{filt}}}}              
\newcommand{\Vcore}{{\Vpsf^{\Tag{core}}}}               
\newcommand{\Vwing}{{\Vpsf^{\Tag{wing}}}}               
\newcommand{\Wfov}{{\V{w}^{\Tag{fov}}}}                 
\newcommand{\Whal}{{\V{w}^{\Tag{hal}}}}                 

\newcommand{\objth}{{o^{\Tag{thr}}}}                    
\newcommand{\medfilt}{{f^{\Tag{med}}}}                  
\newcommand{\J}{\Tag{J}_{2}}                        
\newcommand{\alphap}{\alpha_\Tag{p}}                
\newcommand{\deltap}{\delta_\Tag{p}}                
\newcommand{\lambdap}{\lambda_\Tag{p}}              
\newcommand{\betap}{\beta_\Tag{p}}                      
\newcommand{\Chired}{\chi_{\nu}^{2}}                
\newcommand{\medval}{\includegraphics[height={\f@size pt*2/3}]{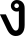}} 
\newcommand{\sym}[1]{$\text{#1}^{\star}$\xspace}

\usepackage{siunitx}
\newcommand{\mas}[1]{\SI{#1}{mas}} 

\newcommand{\ie}{i.e.\xspace}
\newcommand{\eg}{e.g.\xspace}
\newcommand{\vs}{v.s.\xspace}
\newcommand{\st}{\text{\xspace s.t. \xspace}} 

\newcommand{\ElektraFull}{(130)~Elektra\xspace}
\newcommand{\SoneFull}{S/2003 (130) ~1\xspace}
\newcommand{\StwoFull}{S/2014 (130)~1\xspace}
\newcommand{\SthreeFull}{S/2014 (130)~2\xspace}

\newcommand{\Elektra}{Elektra\xspace}
\newcommand{\Sone}{S1\xspace}
\newcommand{\Stwo}{S2\xspace}
\newcommand{\Sthree}{S3\xspace}

\begin{document} 


\title{First observation of a quadruple asteroid}

\subtitle{Detection of a third moon around \ElektraFull  with SPHERE/IFS\thanks{Based on publicly available archival data provided by the ESO Science Archive Facility under programme ID 60.A-9362(A) (Yang et al.)}}

\author{Anthony Berdeu\inst{1,2}
        \and
        Maud Langlois\inst{3}
        \and
        Frédéric Vachier\inst{4}
}

\institute{
        National Astronomical Research Institute of Thailand, 260 Moo 4, T. Donkaew, A. Maerim, Chiang Mai 50180, Thailand
        \and
        Department of Physics, Faculty of Science, Chulalongkorn University,
254 Phayathai Road, Pathumwan, Bangkok 10330, Thailand
        \and
        Université de Lyon, Université Lyon1, ENS de Lyon, CNRS, Centre de Recherche Astrophysique de Lyon UMR 5574, F-69230, Saint-Genis-Laval, France
        \and
        Institut de Mécanique Céleste et de Calcul des Ephémérides, CNRS, Observatoire de Paris, PSL Université, Sorbonne Université, Paris, France
        \\
        \email{anthony@narit.or.th}
        }

\date{Received November 09, 2021 / Accepted January 06, 2022}

 
  \abstract
   {
   Extreme adaptive optics systems, such as the Spectro-Polarimetric High-contrast Exoplanet REsearch facility (SPHERE), push forward the limits in high contrast and high resolution in direct imaging. The main objectives of these instruments are exoplanet detection and characterisation.
    }
   {We aim to increase the contrast limits to detect new satellites orbiting known asteroids. We use cutting-edge data reduction techniques and data processing algorithms that are essential to best analyse the raw data  provided by the instruments and increase their performances. Doing so, the unequalled performances of SPHERE also make it a unique tool to resolve and study asteroids in the solar system, expanding the domain of its main science targets.
   }
   {
   We applied a newly developed data reduction pipeline for integral field spectrographs on archival SPHERE data of a resolved asteroid, \ElektraFull. It was coupled with a dedicated point spread function reconstruction algorithm to model the asteroid halo. Following the halo removal, the moon signal could be extracted more accurately. The moon positions were fitted at three epochs and were used to derive the orbital parameters via a genetic-based algorithm.
  }
   {We announce the discovery of \SthreeFull, a third moon orbiting \ElektraFull, making it the first quadruple asteroid ever found. It is identified in three different epochs, 9, 30, and 31 Dec 2014, at a respective angular separation of \mas{258} (\SI{333}{\kilo\meter}), \mas{229} (\SI{327}{\kilo\meter}), and \mas{319} (\SI{457}{\kilo\meter}). We estimate that this moon has a period of $0.679 \pm 0.001$ day and a semi-major axis of $344 \pm \SI{5}{\kilo\meter}$, with an eccentricity of $0.33 \pm 0.05$ and an inclination of $38 \degr \pm 19 \degr$ compared to the primary rotation axis. With a relative magnitude to the primary of $10.5\pm0.5$, its size is estimated to be $1.6 \pm \SI{0.4}{km}$.
   }
   {
   The orbital parameters of \SthreeFull are poorly constrained due to the unfavourable configurations of the available fragmentary data. Additional observations are needed to better estimate its orbit and to suggest a formation model. This new detection nonetheless shows that dedicated data reduction and processing algorithms modelling the physics of the instruments can push their contrast limits further.
   }


   \maketitle
%

\let\oldpageref\pageref
\renewcommand{\pageref}{\oldpageref*}





\section{Introduction}

More than two decades ago, the advent of adaptive optics (AO) systems triggered a leap forward in the study of asteroids. The first satellite imaged, aside from a spacecraft flyby, was that of (45) Eugenia \citep{Merline:99_45Eugenia} using one of the first (classical) AO systems applied to asteroids (Probing the Universe with Enhanced Optics, PUEO, at the Canada-France-Hawaii Telescope). At the same time, other methods \citep[see][for a complete summary of the early methods]{Merline:02_book} were also closing the gap to detect satellites: light curve methods, radar, Hubble Space Telescope imaging, and direct imaging without AO from the ground, and even occultations.

During the last decade, extreme AO (XAO) systems have pushed the instruments to even higher contrast and resolution \citep{Jovanovic:15, Fusco:16_SAXO}. Such systems, developed for exoplanet research, are also being applied to asteroids \citep{Marchis:14_AO}. It is clear that these XAO systems will be required to get the best data on exoplanets and asteroids from existing and newer, larger telescopes.

\textbf{High resolution -- } The Zurich Imaging Polarimeter instrument \citep[ZIMPOL,][]{Schmid:18_ZIMPOL} on the Spectro-Polarimetric High-contrast Exoplanet REsearch adaptive optics system and coronagraphic facility \citep[SPHERE,][]{Beuzit:19_SPHERE}, on the Very Large Telescope (VLT), works in the visible band ($500$ to $\SI{900}{\nano\meter}$) and possesses an unequalled angular resolution. Resolving the asteroid surfaces paves the way to their topology analysis and their 3D shape estimation \citep{Marchis:21_Kleopatra,Vernazza:21_survey_asteroid}. This gives access to their density and provides some hints as to their composition and origin.

\textbf{High contrast -- } Reaching deeper contrasts naturally led to the detection of faint moons orbiting asteroids, such as in the system of (87)~Sylvia, the first trinary asteroid, discovered by \cite{Marchis:05_87Sylvia} with the NACO (NAOS - CONICA, Nasmyth Adaptive Optics System - Near-Infrared Imager and Spectrograph) instrument on the VLT. Resolving and monitoring moon systems are keys to probing the gravitational field of these complex objects and understanding their dynamics \citep[see \eg][]{Berthier:14_sylvia,Pajuelo:18_107Camilla,Carry:21_sylvia,Marchis:21_Kleopatra}.

In addition to ZIMPOL, SPHERE is also equipped with an integral field spectrograph \citep[IFS,][]{Claudi:08_SPHERE_IFS} and the InfraRed Dual-band Imager and Spectrograph \citep[IRDIS,][]{Dohlen:08_IRDIS} working in near-infrared bands (NIR). The primary goals of these instruments are the imaging and the characterisation of exoplanets by achieving higher contrasts via the use of a coronagraphic mask \citep{Vigan:21_SHINE}. However, these instruments can also be used for the observation of extended objects \citep[see \eg][]{Hanus:17_SPHERE_IFS_asteroid, King:19_SPHERE_IFS_Europa, Souami:21_Neptune}. Their better AO and contrast performances at these longer wavelengths compared to ZIMPOL also make them very efficient for the detection of faint moons orbiting close to a bright primary such as the second moon of (107)~Camilla, discovered by \cite{Marsset:16_107Camilla} in IFS data.

Along with this new instrumentation, new data reduction techniques must keep pace. In this Letter, we apply our newly developed reduction pipeline for SPHERE/IFS, PIC \cite[Projection, Interpolation, Convolution,][]{Berdeu:20_PIC} on an archival dataset of asteroid \ElektraFull (hereafter \Elektra). By correcting the artefacts of the SPHERE Data Centre pipeline \citep[SPHERE/DC,][]{Delorme:17_SPHERE_pipeline}, PIC strongly improves the quality of the data reductions. After the careful removal of the asteroid halo with our new dedicated point spread function (PSF) reconstruction algorithm, we discovered and extracted a new satellite orbiting \Elektra, making it the third satellite in this system, designated as \SthreeFull (hereafter \Sthree). Its first satellite, \SoneFull (hereafter \Sone), was discovered by \cite{Merline:03_130S1} using the Keck AO system and its second satellite, \StwoFull (hereafter \Stwo), was discovered by \cite{Yang:16_Elektra_Minerva} jointly in IFS and IRDIS data, in the same dataset that we use in this Letter.

\Elektra is the first quadruple system ever detected. A preliminary dynamical study is performed on these fragmentary data to confirm that this detection is not an artefact. Additional observations are needed to better constrain the orbit of \Sthree and propose a formation model.

\section{SPHERE/IFS observation and data reduction}

\Elektra was observed on 6 Dec 2014 and 9 Dec 2014 as part of the SPHERE science verification program by Yang et al. (60.A-9362(A) -- `Origin of Multiple Asteroid Systems by Component-Resolved Spectroscopy'). They obtained additional observations on 30 Dec 2014 and 31 Dec 2014, using director's discretionary time \citep{Yang:16_Elektra_Minerva}. These data were acquired using the IFS of SPHERE via its YJH filter ($0.95$ to $\SI{1.65}{\micro\meter}$). We retrieved these publicly available datasets from the European Southern Observatory (ESO) archives. Nonetheless, we could not find the IFS data of 6 Dec 2014 and they are consequently not presented in this Letter\footnote{On this matter, we contacted the ESO data archive facility that could not recover this dataset. At the time of submission, it seems that this IFS dataset is unavailable.}.

The raw data were reduced using the PIC pipeline \citep{Berdeu:20_PIC}. The reduction is based on an inverse problem approach solved via a robust penalisation with spectral and spatial regularisations. To show the relevance of the PIC pipeline, the dataset of 9 Dec 2014 was also reduced with the SPHERE/DC  \citep{Delorme:17_SPHERE_pipeline}. The two pipelines are compared in detail in \refapp{app:DC}.

We would like to mention here that data were simultaneously acquired with IRDIS \citep{Yang:16_Elektra_Minerva}, but they are highly noisy due to inappropriate detector integration time (DIT) selection and we consequently could not detect the presence of \Sthree in these datasets. As a consequence, they are not presented in this Letter.

\section{Image processing: Overview}
\label{sec:IP_overview}

Angularly speaking, the two already known moons around \Elektra revolve very closely to the asteroid and \Stwo is nearly buried in its halo \citep{Yang:16_Elektra_Minerva}. The detection of an even fainter and closer moon implies a careful estimate and removal of this halo.

For the detection of \Stwo, \cite{Yang:16_Elektra_Minerva} straightforwardly adapted a technique described by \cite{Wahhaj:13_GPI_arc}. Initially developed for exoplanet detection to remove residual speckles in coronagraph images, it uses local medians on arcs centred on the primary, removed from each pixel. This technique does not account for the physics of the halo that is poorly estimated, and it can result in some self-subtraction that can bias the moon photometry or even prevent its detection.

To circumvent these effects, we present a halo removal technique here based on a physical model to describe its shape. The halo comes from the faint extensions, so-called wings, of the PSF that make the light of the asteroid scatter far from its photometric surface. In other words, the halo can be modelled by the convolution of the asteroid image with the wings of the instrument PSF. But the sharp image of the asteroid is unknown and the PSF cannot be predicted as it depends on the seeing conditions and the AO performances during each acquisition. As a consequence, these two components of the model must be estimated jointly and directly from the data, a problem known as blind deconvolution \citep{Lam:00,Mugnier:04_MISTRAL,Soulez:12}.

\begin{figure}[ht!]
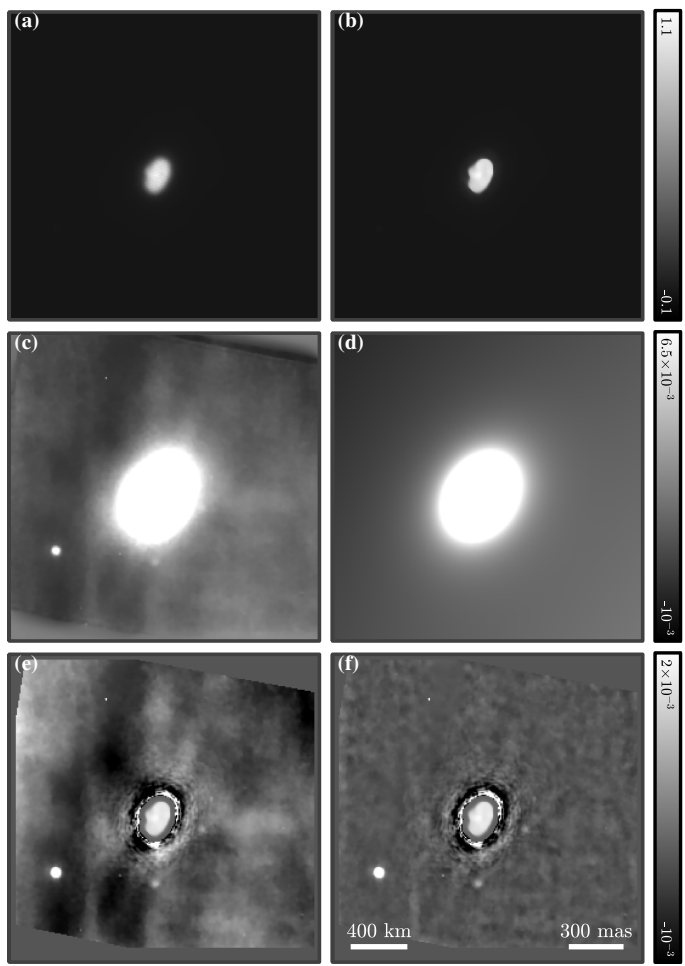
 
        \centering
        
        \newcommand{\LineRatio}{1}
        
        \newcommand{\PathFig}{IP_overview_}
        \newcommand{\FlagReduc}{_PIC}
        \newcommand{\FigOne}{\PathFig Step1_raw_data_obj\FlagReduc .pdf}
        \newcommand{\FigTwo}{\PathFig Step2_deconvolution_obj\FlagReduc .pdf}
        \newcommand{\FigThree}{\PathFig Step1_raw_data_obj_bar.pdf}
        \newcommand{\subfigColor}{white}        
        
        \sbox1{\includegraphics{\FigOne}}               
        \sbox2{\includegraphics{\FigTwo}}               
        \sbox3{\includegraphics{\FigThree}}     
        \newcommand{\ColumnWidth}[1]
                {\dimexpr \LineRatio \linewidth * \AspectRatio{#1} / (\AspectRatio{1} + \AspectRatio{2} + \AspectRatio{3}) \relax
                }
        \newcommand{\ColumnGap}{\hspace {\dimexpr \linewidth /4 - \LineRatio\linewidth /4 }}

        \begin{tabular}{
                @{\ColumnGap}
                C{\ColumnWidth{1}}
                @{\ColumnGap}
                C{\ColumnWidth{2}}
                @{\ColumnGap}
                C{\ColumnWidth{3}}
                @{\ColumnGap}
                }
                \subfigimg[width=\linewidth,pos=ul,font=\fontfig{\subfigColor}]{$\;$(a)}{0.0}{\FigOne} &
                \subfigimg[width=\linewidth,pos=ul,font=\fontfig{\subfigColor}]{$\;$(b)}{0.0}{\FigTwo} &
                \subfigimg[width=\linewidth,pos=ul,font=\fontfig{\subfigColor}]{}{0.0}{\FigThree}
        \end{tabular}
        
        \vspace{-5pt}
        
        \renewcommand{\FigOne}{\PathFig Step1_raw_data_halo\FlagReduc .pdf}
        \renewcommand{\FigTwo}{\PathFig Step3_fitting_halo_mod\FlagReduc .pdf}
        \renewcommand{\FigThree}{\PathFig Step1_raw_data_halo_bar.pdf}
        
        \sbox1{\includegraphics{\FigOne}}               
        \sbox2{\includegraphics{\FigTwo}}               
        \sbox3{\includegraphics{\FigThree}}     

        \begin{tabular}{
                @{\ColumnGap}
                C{\ColumnWidth{1}}
                @{\ColumnGap}
                C{\ColumnWidth{2}}
                @{\ColumnGap}
                C{\ColumnWidth{3}}
                @{\ColumnGap}
                }
                \subfigimg[width=\linewidth,pos=ul,font=\fontfig{\subfigColor}]{$\;$(c)}{0.0}{\FigOne} &
                \subfigimg[width=\linewidth,pos=ul,font=\fontfig{\subfigColor}]{$\;$(d)}{0.0}{\FigTwo} &
                \subfigimg[width=\linewidth,pos=ul,font=\fontfig{\subfigColor}]{}{0.0}{\FigThree}
        \end{tabular}
        
        \vspace{-5pt}
        
        \renewcommand{\FigOne}{\PathFig Step3_fitting_halo_res\FlagReduc .pdf}
        \renewcommand{\FigTwo}{\PathFig Step4_annulus_median\FlagReduc .pdf}
        \renewcommand{\FigThree}{\PathFig Step3_fitting_halo_res_bar.pdf}
        
        \sbox1{\includegraphics{\FigOne}}               
        \sbox2{\includegraphics{\FigTwo}}               
        \sbox3{\includegraphics{\FigThree}}     

        \begin{tabular}{
                @{\ColumnGap}
                C{\ColumnWidth{1}}
                @{\ColumnGap}
                C{\ColumnWidth{2}}
                @{\ColumnGap}
                C{\ColumnWidth{3}}
                @{\ColumnGap}
                }
                \subfigimg[width=\linewidth,pos=ul,font=\fontfig{\subfigColor}]{$\;$(e)}{0.0}{\FigOne} &
                \subfigimg[width=\linewidth,pos=ul,font=\fontfig{\subfigColor}]{$\;$(f)}{0.0}{\FigTwo} &
                \subfigimg[width=\linewidth,pos=ul,font=\fontfig{\subfigColor}]{}{0.0}{\FigThree}
        \end{tabular}
        \caption{\label{fig:IP_overview} Overview of the image processing steps, applied on the $24^{\text{th}}$ acquisition of 9 Dec 2014 after its reduction by PIC \citep{Berdeu:20_PIC}. (a)~Normalised reduced data. (b)~Deconvolved object. (c)~Saturated view of (a). (d)~Saturated view of the fitted halo model. (e)~Residuals after the subtraction from (a,c) of the halo model (d). (f)~Residuals after the application of the annulus median filter on (e). In (e) and (f), the unusable pixels for the halo fitting and the median filtering (edges of the IFS field-of-view and deconvolved primary) have been replaced by a dimmer view of (b) to simultaneously visualise the primary and the moons.}
\end{figure}

Thus, along with the object deconvolution, one objective of the algorithm is to reconstruct the AO-corrected PSF of each reduced hypercube. Our PSF model is based on the sum of two components: its core and its wings. It is similar to what has been proposed by \cite{Fetick:19_PSF_recons} where the PSF is modelled by the convolution of the diffraction-limited PSF of the instrument (shaping its core) by the PSF of the atmosphere fitted with a modified two-dimensional (2D) Moffat function (shaping its wings). The steps of Algorithm~\ref{alg:IP_overview} are detailed in \refapp{app:IP_details} and summarised in \fig{fig:IP_overview}. In short, the core of the PSF was fitted on the brightest moon, see \subfig{fig:IP_deconv}{a}. By deconvolving the reduced image (\subfigs{fig:IP_overview}{a,c}), with this PSF, a sharp image of the primary was obtained, see \subfig{fig:IP_overview}{b}. This sharp image was used to fit the PSF wings and produce a halo model (\subfig{fig:IP_overview}{d}). This model was then removed from the data (\subfig{fig:IP_overview}{e}), that is to say further cleaned with a median filter (\subfig{fig:IP_overview}{f}), to reveal faint objects in the primary vicinity.

The scale bars of \fig{fig:IP_overview} emphasise the importance of the halo removal. Indeed, the intensity on the newly discovered moon in \subfig{fig:IP_overview}{f} is~$\sim 1500$ fainter than the primary surface intensity ($7\times10^{-4}$), while the halo is only~$\sim 150$ fainter ($6\times10^{-3}$) in \subfigs{fig:IP_overview}{a,c}.

\section{Satellites' relative astrometry and photometry}
\label{sec:Pos_fitting}

\begin{figure*}[ht!]
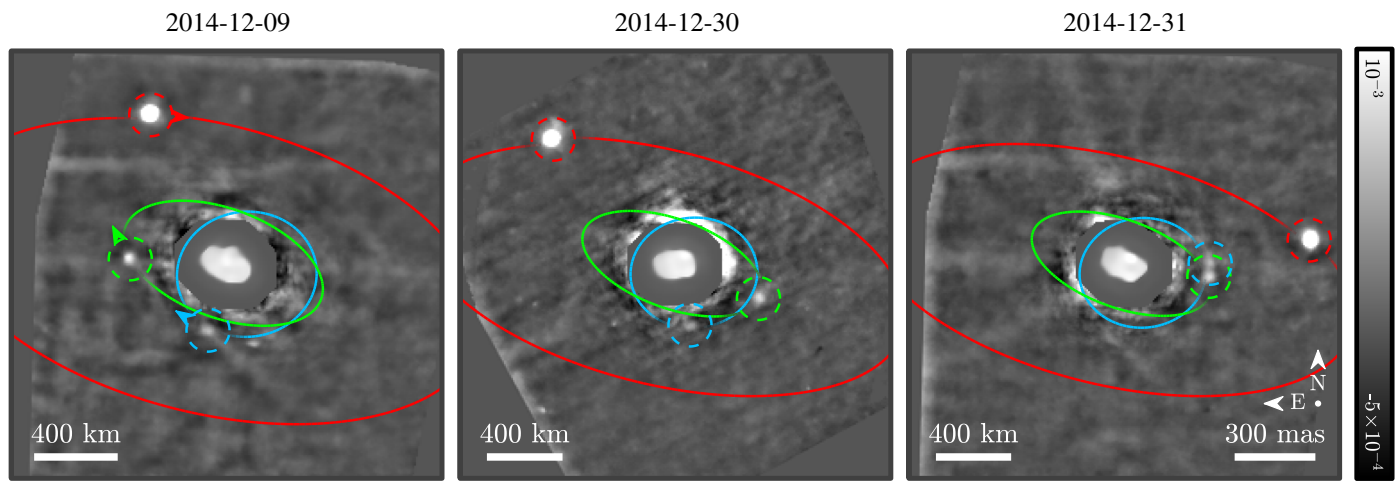
 
        \centering

        \centering
        
        \newcommand{\LineRatio}{1}
        
        \newcommand{\PathFig}{Med_projection_}
        \newcommand{\FlagReduc}{_PIC}
        \newcommand{\FigOne}{\PathFig Elektra_2014-12-09\FlagReduc .pdf}
        \newcommand{\FigTwo}{\PathFig Elektra_2014-12-30\FlagReduc .pdf}
        \newcommand{\FigThree}{\PathFig Elektra_2014-12-31\FlagReduc.pdf}
        \newcommand{\FigFour}{\PathFig Color_bar.pdf}
        \newcommand{\subfigColor}{white}        
        
        \sbox1{\includegraphics{\FigOne}}               
        \sbox2{\includegraphics{\FigTwo}}               
        \sbox3{\includegraphics{\FigThree}}     
        \sbox4{\includegraphics{\FigFour}}              
        \newcommand{\ColumnWidth}[1]
                {\dimexpr \LineRatio \linewidth * \AspectRatio{#1} / (\AspectRatio{1} + \AspectRatio{2} + \AspectRatio{3} + \AspectRatio{4}) \relax
                }
        \newcommand{\ColumnGap}{\hspace {\dimexpr \linewidth /5 - \LineRatio\linewidth /5 }}

        \begin{tabular}{
                @{\ColumnGap}
                C{\ColumnWidth{1}}
                @{\ColumnGap}
                C{\ColumnWidth{2}}
                @{\ColumnGap}
                C{\ColumnWidth{3}}
                @{\ColumnGap}
                C{\ColumnWidth{4}}
                @{\ColumnGap}
                C{\ColumnWidth{5}}
                @{\ColumnGap}
                }
                2014-12-09 &
                2014-12-30 &
                2014-12-31 
                \\
                \subfigimg[width=\linewidth,pos=ul,font=\fontfig{\subfigColor}]{}{0.0}{\FigOne} &
                \subfigimg[width=\linewidth,pos=ul,font=\fontfig{\subfigColor}]{}{0.0}{\FigTwo} &
                \subfigimg[width=\linewidth,pos=ul,font=\fontfig{\subfigColor}]{}{0.0}{\FigThree} &
                \subfigimg[width=\linewidth,pos=ul,font=\fontfig{\subfigColor}]{}{0.0}{\FigFour}
        \end{tabular}
        
        \caption{\label{fig:Orb_median} Median projection of the temporal cubes for the different dates reduced with the PIC pipeline \citep{Berdeu:20_PIC}. The fitted orbits are projected on the different dates: \Sone in red, \Stwo in green, and \Sthree in blue. The arrows indicate the projected motion direction. The dashed circles indicate the median fitted moon position during the acquisitions at each date. The central part of the field of view has been replaced by a dimmer deconvolved image of \Elektra to simultaneously visualise the primary and the moons.}
\end{figure*}

After the image processing described in the previous section, the different frames were centred on the photocentre of \Elektra by computing the centroid of its segmented deconvolved reconstruction and orientated towards the north using calibrations from \cite{Maire:21_astrometry}. The temporal median projections of each sequence are given in \fig{fig:Orb_median}. The three moons are clearly visible at all epochs.

To extract the astrophotometry of the three satellites, we used MPFIT2DPEAK\footnote{Craig B. Markwardt, NASA/GSFC Code 662, Greenbelt, MD 20770, \url{http://cow.physics.wisc.edu/~craigm/idl/idl.html}}, a 2D non-linear least squares Gaussian fitter. It returns the peak intensity, half-widths (independent semi-major and semi-minor axes), and position of each moon for each image of the observation sequences. We included a rejection criteria based on the full-width at half-maximum if greater or smaller by $\SI{60}{\percent}$ from the median (\mas{25}). To account for the satellite motion and to reduce astrometric uncertainty, we refined the astrometry with a temporal linear fit on the polar positions for each epoch, see \fig{fig:DCvsPIC}. {\textcolor{blue}{Visualisations}} {\textcolor{blue}{1}}, {\textcolor{blue}{2,}} and {\textcolor{blue}{3}} show the time-lapses of the linearly fitted positions at the different epochs. The astrophotometry fits on the three moons are listed in the tables of \refapp{app:astrometry}. On average, \Sthree was found in the three different epochs (9, 30 , and 31 Dec 2014) at a respective angular separation of \mas{258} (\SI{333}{\kilo\meter}), \mas{229} (\SI{327}{\kilo\meter}), and \mas{319} (\SI{457}{\kilo\meter}) from \Elektra.

For the pixel scale of the SPHERE/DC pipeline, we used a constant value of $7.46\pm0.02$~mas/pixel \citep{Maire:21_astrometry}. It corresponds to a pixel scale of $6.66\pm0.02$~mas/pixel for the reductions with PIC \citep{Berdeu:20_PIC}. The true north (TN) correction of \SI{-1.75}{\degr} typically has an error of $\pm \SI{0.07}{\degr}$ \citep{Maire:21_astrometry}. This TN error includes the observed systematic error in the parallactic angle estimation due to backlash in the derotator mechanism ($\sim \SI{0.05}{\degr}$), as reported by \cite{Beuzit:19_SPHERE}. For the IFS, an additional offset of \SI{100.48}{\degr} in the clockwise direction was applied to account for the orientation of the instrument's field of view. The astrometric errors were estimated by including the following systematics described by \cite{Langlois:21_SHINE} and \cite{Maire:21_astrometry}: plate scale and TN errors, the error on the satellite centring (estimated from the standard deviation from the linear fit), and the error on the primary centring. The accurate time{\-}stamps were computed from the data \texttt{fits} header for each DIT by including a parametric model of the overheads occurring during the data recording \citep{Delorme:17_SPHERE_pipeline}.

The relative photometry between \Elektra and \Sone was obtained with the ratio of their integrated flux. The integrated flux of \Elektra was calculated by integrating its segmented deconvolved reconstruction. By definition, this measurement does not account for the energy ratio diluted in the halo and attributable to the wing of the PSF. The integrated flux of the brightest moon \Sone was obtained by integrating the 2D Gaussian function fitted for the PSF core fitting (see \refapp{app:IP_core}). Once again, this does not account for the ratio of the flux in the PSF wing (see \subfig{fig:IP_halo}{c}). These techniques make the two measurements coherent and conserve their intensity ratio. The relative photometry between the three moons was obtained by scaling the ratio of their Gaussian amplitude fitted as described above. Based on these measurements, the mean magnitude difference through the YJH band (0.95 to \SI{1.65}{\micro\meter}) between \Elektra and its companions is $7.6\pm0.2$ for \Sone, $10.0\pm0.4$ for \Stwo, and  $10.5\pm0.5$ for \Sthree based on the 9 Dec 2014 observation (see \refapp{app:astrometry}).

Assuming that the moons have the same albedo as the primary, it is possible to estimate their diameter from these relative magnitude. Taking $199\pm\SI{7}{\km}$ as the effective diameter for \Elektra \citep{Hanus:17_SPHERE_IFS_asteroid, Miles:18_occultation}, \Sone is $6.0 \pm \SI{0.6}{km}$, \Stwo is $2.0 \pm \SI{0.4}{km,}$ and \Sthree is $1.6 \pm \SI{0.4}{km}$. Our estimates on \Sone and \Stwo are in agreement with the values announced by \cite{Yang:16_Elektra_Minerva} of $6.0 \pm \SI{1.5}{km}$ and $2.0 \pm \SI{1.5}{km}$, respectively. In addition, the size of \Sone was estimated to be $\SI{4}{km}$ by \cite{Merline:03_130S1} and $\SI{7}{km}$ by \cite{Marchis:08}.

\section{Orbital fitting}
\label{sec:Orb_fitting}

The gravitational environment, in which \Sthree evolves, is dominated by the zonal coefficient $\J$ \citep{Yang:16_Elektra_Minerva}. The observations of \Sthree that we measured are separated only by 3 weeks, which represents about 33 revolutions. This is sufficient for the satellite to be disturbed by the $\J$. However, the position of \Sthree  in the observations are all located on the same side of the orbit, as is visible in \fig{fig:Orb_median}. 
Thus, the distribution of the observations is fragmentary, both spatially and temporally. This is an unfavourable situation in which to constrain the orbit parameters. Moreover, taking the $\J$ into account is certainly insufficient to describe the gravitational field generated by \Elektra over a long period of time.

As a consequence, we fitted Keplerian orbits for the three moons ($\J=0$), centred on \Elektra's photocentre. Such Keplerian model is generally more robust and stable in such a situation \citep{Yang:16_Elektra_Minerva}.

We used the Genoid algorithm \citep[GENetic Orbit IDentification,][]{Vachier:12_Genoide,Berthier:14_sylvia,Pajuelo:18_107Camilla,Carry:19_daphne} to fit the Keplerian model to the observations. It is a genetic-based algorithm that relies on a meta-heuristic method to find the best-fit (\ie minimum $\chi^{2}$) set of dynamical parameters (among others: mass $M$, period $P$, semi-major axis $a$, eccentricity $e$, inclination $i$, longitude of the node $\Omega$, argument of pericentre $\omega$, and time of passage to pericentre $t_\Tag{p}$) by refining, generation after generation, a grid of test values.

For a purely Keplerian fit, the unknowns driving the system dynamics are the system mass and the six independent orbital parameters of each moon listed above ($M$, $P$, and $a$ are linked according to Kepler's second law). Standard detection techniques combine the frames in each epoch to increase the signal-to-noise ratio, for example by averaging the frames or taking their median. This leads to only two data points per night ($x$ and $y$ positions), and thus here six points in total for each moon. This would make the system under-constrained, missing one independent measurement. With our method, the efficiency of the halo removal allows one to fit the astrometry of the moons in almost each frame of the different epochs. Combined with the robustness of the linear fit, it is then possible to fit a small arc of the orbit in each dataset (see \fig{fig:DCvsPIC}). If all these measurements are not independent, they at least give access to the local derivatives of each moon orbit, increasing the number of data points per night to four ($x$ and $y$ positions and their derivatives). These derivatives help Genoid to reject orbits passing through the data points, but with the wrong local tangent. It is then possible to fit all the parameters of the Keplerian model. Nonetheless, some parameters will still be poorly constrained, as shown below.

As detailed in \refapp{app:Orbit_fitting_pole}, additional archival data on \Sone are available, ranging from 2003 to 2019. These data are used to estimate a first guess on the system mass and \Sone{}'s orbital parameters. This guess is used to initialise the Genoid fit on the 2014 datasets where \Elektra's mass and the orbital parameters of the three moons are then let free to evolve. As further discussed in \refapp{app:Orbit_fitting_pole}, the evolution of the viewing angle on the system between 6 Dec and 31 Dec, combined with the knowledge of the local arcs of the orbits with a good astrometry precision, are sufficient for Genoid to favour one orbital pole solution of its different degenerate symmetries. We present this solution in the following.

This global fit provides an overall root mean square (RMS) error of $\mas{6.8}$ between the positions predicted by the model and the astrometry measurements. This error is in good agreement with the uncertainties on the astrometry, listed in \refapp{app:astrometry}. The fit RMS error per moon is  $\mas{3.5}$ for \Sone, $\mas{2.4}$ for \Stwo, and $\mas{11.3}$ for \Sthree. This last value can suggest that a Keplerian fit is only a first order approximation. It is also consistent with the fact that despite the halo removal, the residuals are still high and possibly corrupt the astrometry fit, as shown by the dispersion of the measurements in \fig{fig:DCvsPIC}.

\Elektra's estimated derived mass is $\Paren{7.0 \pm 0.3} \times 10^{18} \, \SI{}{\kilo\gram}$. The orbital elements of \Sthree are given in \tab{tab:Orb_fitting_S3}. These values should be considered with great precaution. The orbit seems very eccentric ($e=0.33 \pm 0.05$) and inclined with respect to the spin axis of \Elektra ($38 \degr \pm 19 \degr$), as shown in \tab{tab:Spin_fitting} of \refapp{app:Orbit_fitting_spin}. Also, the orbital elements of \Sone and \Stwo are listed in \refapp{app:Orbit_fitting_S1S2}. 

\begin{table}
\begin{center}
  \caption{Orbital elements of \Sthree expressed in EQJ2000.}
    \label{tab:Orb_fitting_S3} 
  
    \Sthree{} -- \SthreeFull
    \vspace{5pt}
  
  \begin{tabular}{l ll ll}
    \hline\hline
    \noalign{\smallskip}
    \multicolumn{2}{c}{Observing data set} \\
    \noalign{\smallskip}
    Number of observations & 120   \\
    Time span (days)       & 22 &    \\
    RMS (mas)              & 11.31   \\
    \hline
    \noalign{\smallskip}
    \multicolumn{2}{c}{Orbital elements EQJ2000} \\
    \noalign{\smallskip}
    $P$ (day)        &       0.679   & $\pm$ 0.001  \\
    $a$ (km)         &     344       & $\pm$ 5       \\
    $e$              &       0.33    & $\pm$ 0.05    \\
    $i$ (\degr)      &     129       & $\pm$ 24      \\
    $\Omega$ (\degr) &     127       & $\pm$ 18      \\
    $\omega$ (\degr) &      23       & $\pm$ 11      \\
    $t_\Tag{p}$ (JD)     & 2456990.51    & $\pm$ 0.03    \\
    $\paren{\alphap,\deltap}$ (\degr) & $\Paren{37, -40}$ & $\Paren{\pm 18, \pm 24}$       \\
    $\paren{\lambdap,\betap}$ (\degr) & $\Paren{17, -50}$ & $\Paren{\pm 28, \pm 20}$       \\
    \hline
  \end{tabular}
  \tablefoot{Orbital period $P$, semi-major axis $a$, eccentricity $e$, inclination $i$, longitude of the ascending node $\Omega$, argument of pericentre $\omega$, time of pericentre $t_\Tag{p}$, ecliptic J2000 coordinates of the orbital pole $\paren{\lambdap,\betap}$, and equatorial J2000 coordinates of the orbital pole $\paren{\alphap,\deltap}$. The number of observations and RMS between the predicted and observed positions are also provided. Uncertainties are given at 1-$\sigma$.}
\end{center}
\end{table}

Figure \ref{fig:Orb_median} shows the orbits of the different satellites. \Sthree is highlighted in blue. For the observation of 31 Dec 2014, we can see a close projected approach of \Sthree with \Stwo, at more than \SI{400}{\kilo\meter} from \Elektra. This emphasises the high eccentricity of the orbit.

In \refapp{app:Extra_dates}, we present some hints for the presence of \Sthree in other datasets obtained with the IFS in 2016 and ZIMPOL in 2019. They are nonetheless temporally too far from the 2014 observations to be included in the orbit fitting presented in this section. Taking them into account requires substantial work that is beyond the scope of this Letter.

\section{Discussion and conclusion}

In this Letter, we presented the discovery of \Sthree, a third moon orbiting \Elektra. The newly detected satellite revolves inside the orbit of \Stwo with a semi-major axis of $344 \pm \SI{5}{km}$ and an orbital period of $0.679 \pm 0.001$ days around the primary.  Nonetheless, a lot of uncertainties remain concerning the orbit of \Sthree. More data on \Stwo and \Sthree, as well as a more thorough dynamical study are necessary to solve the problem of the motion of the satellites of \Elektra. However, the discovery of the first quadruple asteroid system slightly opens the way for understanding the mechanisms of the formation of these satellites.

In terms of data processing, \Sthree is barely visible in the data reduced with the standard pipeline and processed with standard halo removal algorithms and it was missed until now. This shows that the development of dedicated inverse problem algorithms based on a forward modelling of the instrument is key to expanding the instrument capacities. The PIC reduction pipeline permits the suppression of artefacts that could hide small moons or lead to false detections. In addition, modelling the instrument PSF allows one to remove the asteroid halo carefully so as to increase the contrast in its surrounding. The method presented in this Letter paves the way for more robust and general approaches to reconstruct more complex PSFs of extended objects corrupted by AO corrections to study their close vicinity.

\begin{acknowledgements}
        The authors would like to warmly thank the Referee Bill Merline for his pertinent feedback and remarks that helped them improving the quality and clarity of this Letter.
        \\
        The authors are thankful to Ferréol Soulez for the fruitful discussions. The authors thank Jules Dallant for his Keplerian orbit visualisation software which initially helped the authors to confirm that the positions fitted on \Sthree were consistent with a Keplerian motion. The authors also thank Anthony Boccaletti for his IDL tool we used to accurately estimate the acquisition timestamps by taking into account the overheads during the data recording.
        \\
        This work has made use of the SPHERE Data Centre, jointly operated by OSUG/IPAG (Grenoble), PYTHEAS/LAM/CeSAM (Marseille), OCA/Lagrange (Nice), Observatoire de Paris/LESIA (Paris), and Observatoire de Lyon/CRAL, and supported by a grant from Labex OSUG@2020, within the program `Investissements d'Avenir' (ANR-10-LABX-56).
        \\
        This work was supported by the Programme National de Planétologie (PNP) of CNRS/INSU, co-funded by CNES. This work was carried out within the LabEx LIO (ANR-10-LABX-0066) of the University of Lyon, created within the framework of the Future Investments Program bearing the reference ANR-10-LABX-0066 implemented placed by the State and managed by the National Research Agency (ANR).
        \\
        All the reductions are based on public data provided by the ESO Science Archive Facility and acquired for different programs: 60.A-9362(A) (Yang et al., `Origin of Multiple Asteroid Systems by Component-Resolved Spectroscopy' -- 9, 30 and 31 Dec 2014),  296.C-5038(A) (Yang et al., `SPHERE Follow-up of the New Triple Asteroid (130) Elektra' -- 16 Feb 2016) and 199.C-0074(E) (Vernazza et al., `Asteroids as tracers of solar system formation: Probing the interior of primordial main belt asteroids' -- 30 Jul 2019 and 6 Aug 2019). For the initial fit of \Sone's orbital parameters, additional archival data are used, as detailed in \refapp{app:Orbit_fitting_pole}.
        \\
        The authors acknowledge the supports from Chulalongkorn University’s CUniverse (CUAASC) grant and from the Program Management Unit for Human Resources \& Institutional Development, Research and Innovation, NXPO (grant number B16F630069).
\end{acknowledgements}

%
%
\bibliographystyle{aa}
\bibliography{bib_Elektra}


\begin{appendix}

\section{Image processing: Halo removal algorithm}
\label{app:IP_details}

In \refsec{sec:IP_overview}, we briefly presented the image processing we developed to remove the halo of \Elektra. It was applied to each acquisition, that is to say on each $\text{2D}+\lambda$ hypercube~$\Vcube$ provided by the IFS reduction pipeline. The main steps, also summed up in the pseudp-code of Algorithm~\ref{alg:IP_overview}, are as follows: $\Vred$, the reduced image, was obtained by averaging the IFS $\text{2D}+\lambda$ hypercube~$\Vcube$ along its spectral dimension and was normalised to 1 (see \subfig{fig:IP_overview}{a}); $\Vcore$, the core part of the PSF, was estimated by fitting a 2D Gaussian function on the brightest moon (see \refapp{app:IP_core}); $\Vdec$, the sharp image of the primary, was obtained by deconvolving $\Vred$ with~$\Vcore$ (see \subfig{fig:IP_overview}{b} and \refapp{app:IP_deconv}); $\Vthr$, the primary image, was segmented by keeping only the pixels above~$\objth>25\%$ (see \refapp{app:IP_halo}); $\Vwing$, the faint PSF extentions, were estimated by deconvolving of the halo by~$\Vthr$ (see \refapp{app:IP_halo}); $\Vhalo$, the model halo, is the convolution of the primary image~$\Vthr$ with the total PSF~$\Vpsf = \Vcore + \Vwing$ (see \subfig{fig:IP_overview}{d}); $\Vres$ are the residuals after the halo model removal in which the moons are visible (see \subfig{fig:IP_overview}{e}); $\Vfilt$ was obtained by cleaning the residual background structures with a median filter (see \subfig{fig:IP_overview}{f} and \refapp{app:IP_median}).

\begin{algorithm}
\caption{\label{alg:IP_overview} Overview of the image processing steps performed on each reduced hypercube~$\Vcube$.}
\begin{algorithmic}[1]
\small

\State $\Vred \underset{\text{\subfig{fig:IP_overview}{a}}}{\gets} \Avg{\Vcube\Paren{\x,\lambda}}_{\lambda}$
        \commentalgo{Spectral projection of the reduced cube}

\State $\Vcore \underset{\text{\subfig{fig:IP_deconv}{b}}}{\gets} \Vmoon \simeq \Vcore$
        \commentalgo{Fitting the PSF core on the brightest moon}

\State $\Vdec \underset{\text{\subfig{fig:IP_overview}{b}}}{\gets} \Vred \simeq \Vcore \conv \Vdec$
        \commentalgo{Deconvolution by the PSF core}
        
\State $\Vthr\Paren{\x} \underset{\text{\subfig{fig:IP_halo}{a}}}{\gets}
        \begin{cases}
                0 \text{ if } \Vdec\Paren{\x} \leq \objth
                \\
                \Vdec\Paren{\x} \text{ otherwise}
        \end{cases}$
        \commentalgo{Threshold on the primary}
        
\State $\Vwing \underset{\text{\subfig{fig:IP_halo}{b}}}{\gets} \Vred \simeq \Paren{\Vcore + \Vwing} \conv \Vthr$
        \commentalgo{Fitting the PSF wings}
        
\State $\Vhalo \underset{\text{\subfig{fig:IP_overview}{d}}}{\gets} \Paren{\Vcore + \Vwing} \conv \Vthr$
        \commentalgo{Model of the halo}
        
\State $\Vres \underset{\text{\subfig{fig:IP_overview}{e}}}{\gets} \Vred-\Vhalo$
        \commentalgo{Residuals after removing the halo model}
        
\State $\Vfilt \underset{\text{\subfig{fig:IP_overview}{f}}}{\gets} \Vres - \medfilt\Paren{\Vres}$
        \commentalgo{Residuals of the median annulus filtering}
\end{algorithmic}
\end{algorithm}

In the following, we describe this algorithm in further detail. The code is implemented in Matlab$^{\mathrm{TM}}$. The linear inverse problems use the open-source GlobalBioIm framework \citep{code:GlobalBioIm_19}.

\subsection{Fitting the PSF core on the brightest moon}
\label{app:IP_core}

\begin{figure}[ht!]
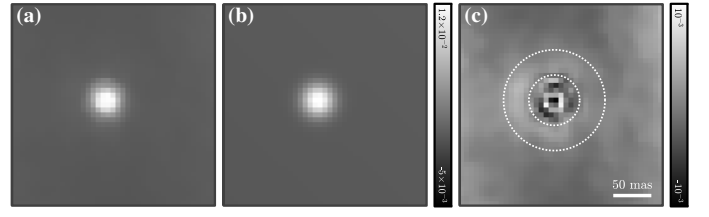
 
        \centering
        
        \newcommand{\LineRatio}{1}
        
        \newcommand{\PathFig}{IP_deconv_}
        \newcommand{\FlagReduc}{_PIC}
        \newcommand{\FigOne}{\PathFig PSF\FlagReduc .pdf}
        \newcommand{\FigTwo}{\PathFig PSF_model\FlagReduc .pdf}
        \newcommand{\FigThree}{\PathFig PSF_bar.pdf}
        \newcommand{\FigFour}{\PathFig PSF_residuals\FlagReduc .pdf}
        \newcommand{\FigFive}{\PathFig PSF_residuals_bar.pdf}
        \newcommand{\subfigColor}{white}        
        
        \sbox1{\includegraphics{\FigOne}}               
        \sbox2{\includegraphics{\FigTwo}}               
        \sbox3{\includegraphics{\FigThree}}     
        \sbox4{\includegraphics{\FigFour}}      
        \sbox5{\includegraphics{\FigFive}}      
        \newcommand{\ColumnWidth}[1]
                {\dimexpr \LineRatio \linewidth * \AspectRatio{#1} / (\AspectRatio{1} + \AspectRatio{2} + \AspectRatio{3} + \AspectRatio{4} + \AspectRatio{5}) \relax
                }
        \newcommand{\ColumnGap}{\hspace {\dimexpr \linewidth /6 - \LineRatio\linewidth /6 }}

        \begin{tabular}{
                @{\ColumnGap}
                C{\ColumnWidth{1}}
                @{\ColumnGap}
                C{\ColumnWidth{2}}
                @{\ColumnGap}
                C{\ColumnWidth{3}}
                @{\ColumnGap}
                C{\ColumnWidth{4}}
                @{\ColumnGap}
                C{\ColumnWidth{5}}
                @{\ColumnGap}
                }
                \subfigimg[width=\linewidth,pos=ul,font=\fontfig{\subfigColor}]{$\,$(a)}{0.0}{\FigOne} &
                \subfigimg[width=\linewidth,pos=ul,font=\fontfig{\subfigColor}]{$\,$(b)}{0.0}{\FigTwo} &
                \subfigimg[width=\linewidth,pos=ul,font=\fontfig{\subfigColor}]{}{0.0}{\FigThree} &
                \subfigimg[width=\linewidth,pos=ul,font=\fontfig{\subfigColor}]{$\,$(c)}{0.0}{\FigFour} &
                \subfigimg[width=\linewidth,pos=ul,font=\fontfig{\subfigColor}]{}{0.0}{\FigFive}
        \end{tabular}   
        
        \caption{\label{fig:IP_deconv} Fitting the core of the PSF on the brightest moon from \subfig{fig:IP_deconv}{c}. (a)~Normalised reduced data~$\Vmoon$. (b)~Fitted PSF model and background~$\Vcore + \Vbg$. (c)~Residuals of the model. For comparison with the PSF size, the edges of the annulus filter are highlighted by the dotted white lines.}
\end{figure}

The image of the asteroid can only be obtained with the deconvolution of~$\Vred$ by the PSF. To get the details of the asteroid surface and its edges, only the core of the PSF~$\Vcore$ is needed. The PSF extensions only influence large-scale structures in the data, such as the halo and thus the photometry, which are not relevant for this step. An estimate of this central part of the PSF was obtained on the brightest moon, shown in \subfig{fig:IP_deconv}{a}, by fitting a 2D Gaussian function 
\begin{equation} 
        \label{eq:PSF_core}
        \Vcore\Paren{\x}
        =
        g\Paren{\x, a, \alpha, \sig}
        \triangleq
        a \E{-0.5\Paren{
                {r_{1}^{2}}/{\sigma_{1}^{2}}
                + {r_{2}^{2}}/{\sigma_{2}^{2}}}}
        \,,
\end{equation} 
where $\x=\Paren{x,y}$ are the 2D spatial coordinates; $a$ is the amplitude of the 2D Gaussian pattern; $\alpha$ is the orientation of the 2D Gaussian pattern; $r_{1} = x\cos{\alpha} + y\sin{\alpha}$ and $r_{2} = -x\sin{\alpha} + y\cos{\alpha}$ are the coordinates in the rotated frame of the Gaussian pattern; and $\sig = \Paren{\sigma_{1}, \sigma_{2}}$ are the standard deviations of the Gaussian pattern along its two axes.

These parameters must be fitted from the data~$\Vmoon$ (see \subfig{fig:IP_deconv}{a}), where an offset background~$\Vbg$ can corrupt their estimation. The background is described as follows:
\begin{equation} 
        \label{eq:BG}
        \Vbg\Paren{\x, \V{c}} = c_{0} + c_{1} x+ c_{2} y
        \,,
\end{equation}
where~$\V{c} = \Paren{c_{0},c_{1},c_{2}}$ is the list of the three coefficients of a one degree polynomial.
In doing so, the data can be written as follows:
\begin{equation}
        \Vmoon\Paren{\x} \simeq g\Paren{\x, a, \alpha, \sig} + \Vbg\Paren{\x, \V{c}}
        \,,
\end{equation}
where~$\simeq$ is to symbolise the uncertainties from the noise.

All of these parameters were estimated in an alternate algorithm. First, the 2D Gaussian parameters were obtained by minimising
\begin{equation} 
        \label{eq:solve_gauss}
        \underset{\x_{0}, a, \alpha, \sig}{\text{argmin}}\;\sum_{\x}\Paren{\Vmoon\Paren{\x} - g\Paren{\x-\x_{0}, a, \alpha, \sig} - \Vbg\Paren{\x, \V{c}} }^{2}
        \,,
\end{equation}
by keeping~$\V{c}$ fixed and where~$\x_{0}$ is the position of the moon. Then, the background parameters were estimated by solving
\begin{equation} 
        \label{eq:solve_bg}
        \underset{\V{c}}{\text{argmin}}\;\sum_{\x}\Abs{\Vmoon\Paren{\x} - g\Paren{\x-\x_{0}, a, \alpha, \sig} - \Vbg\Paren{\x, \V{c}} }
        \,,
\end{equation}
by keeping~$\Paren{\x_{0}, a, \alpha, \sig}$ fixed. Summing the absolute value is a robust method to reduce the impact of outlier pixels in the cost function.

This algorithm is summarised in Algorithm~\ref{alg:IP_core}. The optimisation problem \eq{eq:solve_gauss} is solved with the VMLM-B algorithm \citep[Variable Metric with Limited Memory-Bounded, ][]{Thiebaut:02}, a limited-memory quasi-Newton method with BFGS updates \citep[Broyden-Fletcher-Goldfarb-Shanno, ][]{Nocedal:80} that handles bound constraints. The optimisation problem \eq{eq:solve_gauss} is solved with the simplex search method of \cite{Lagarias:98_fminsearch}.

\begin{algorithm}
\caption{\label{alg:IP_core} Pseudo-code of the PSF core fitting on the brightest moon.}
\begin{algorithmic}[1]
\small

\State $\V{c} \gets \V{0}$
        \commentalgo{Initialisation with no background}

\For{$i\text{ from } 1 \text{ to } 5$}
        \commentalgo{$Five$ iterations of the alternate optimisation}
        
        \State $\Paren{a, \alpha, \sig} \gets $ solving \eq{eq:solve_gauss}
                \commentalgo{PSF core fitting}
        \State $\V{c} \gets $ solving \eq{eq:solve_bg}
                \commentalgo{Background fitting}
                        
\EndFor
\end{algorithmic}
\end{algorithm}

As seen in \subfigs{fig:IP_deconv}{b,c}, this model is sufficient to describe the core of the PSF. \subfigfull{fig:IP_deconv}{c} also shows that the wings of the PSF cannot be fitted on the brightest moon because the signal far from its core is too noisy.

\subsection{Image deconvolution}
\label{app:IP_deconv}

The image~$\Vdec$ of the primary was obtained by deconvolving the reduced data
\begin{equation}
        \Vred \simeq \Vcore \conv \Vdec
,\end{equation}
where~$\simeq$ is to symbolise the uncertainties from the noise. This was done by solving the minimisation problem
\begin{equation} 
        \label{eq:solve_deconv}
        \underset{\Vdec\geq0}{\text{argmin}}\; \mathscr{D}\Paren{\Vred,\Vcore \conv \Vdec,\Wfov}
        +
        \mu \mathscr{R}_{\Tag{2D}}\Paren{\Vdec}
        \,,
\end{equation}
where $\Vdec\geq0$ is a positivity constraint on the reconstruction; $\Wfov$ is the binary mask on the usable pixels of the reduced data~$\Vred$ (1 if inside the IFS field of view, see pixels colored in blue, red, and green in \subfig{fig:IP_halo}{a}), and 0 otherwise; $\mathscr{D}\Paren{\Vred,\Vcore \conv \Vdec,\Wfov}$ is the data fidelity term between the~$\Vred$ and the model~$\Vcore \conv \Vdec$ weighted by~$\Wfov$
\begin{equation}
        \mathscr{D}\Paren{\V{\varphi}_{1},\V{\varphi}_{2},\V{w}}
        \triangleq
        \sum_{\x \st \V{w}\Paren{\x} = 1} \Paren{\V{\varphi}_{1}\Paren{\x} -\V{\varphi\Paren{\x}}_{2}}^{2}
        \,;
\end{equation}
$\mathscr{R}_{\Tag{2D}}\Paren{\Vdec}$ is a regularisation term that favours sharp-edged objects to control the spatial continuity and smoothness of the reconstruction; and $\mu$ is the hyperparameter to weigh the regularisation compared to the data fidelity term.

Enforcing an edge-preserving smoothness via~$\mathscr{R}_{\Tag{2D}}\Paren{\Vdec}$ was done by encouraging the sparsity of spatial gradients \citep{Rudin:92_TV, Charbonnier:97_TV}:
\begin{equation} 
        \label{eq:reg_2D}
        \mathscr{R}_{\Tag{2D}}\left(\V{\varphi}\right)
        \triangleq
        \sum_{\x}\Brack{\sqrt{\Paren{\brack{\boldsymbol\nabla_{x}\V{\varphi}\Paren{\x}}^{2}+\brack{\boldsymbol\nabla_{y}\V{\varphi}\Paren{\x}}^{2}+\epsilon^{2}}}-\epsilon}\,,
\end{equation}
where~$\boldsymbol\nabla_{x}$ and $\boldsymbol\nabla_{y}$ correspond to finite difference operators along the first and second spatial dimensions, respectively, and $\epsilon>0$ is a threshold corresponding to the smallest gradient at a sharp edge and it ensures that this hyperbolic approximation of the Euclidean norm is differentiable at zero.

The optimisation problem \eq{eq:solve_deconv} was solved by the VMLM-B algorithm \citep{Thiebaut:02}. The value of $\epsilon = 3 \times 10^{-2}$ and $\mu = 5 \times 10^{-3}$ were manually tuned to give a sharp image while limiting the flat areas. 

As seen in \subfig{fig:IP_overview}{b}, this deconvolution successfully provides an image~$\Vdec$ of \Elektra with sharp edges with some spatial features on its surface. As the PSF extensions have not been included in the model yet, this reconstruction is still corrupted by the asteroid halo. Thus, this image is segmented to extract the primary~$\Vthr$ from its halo by keeping only the pixels above~$\objth>25\%$, emphasised by the green pixels in \subfig{fig:IP_halo}{a}.

\subsection{Halo fitting}
\label{app:IP_halo}

\begin{figure}[ht!]
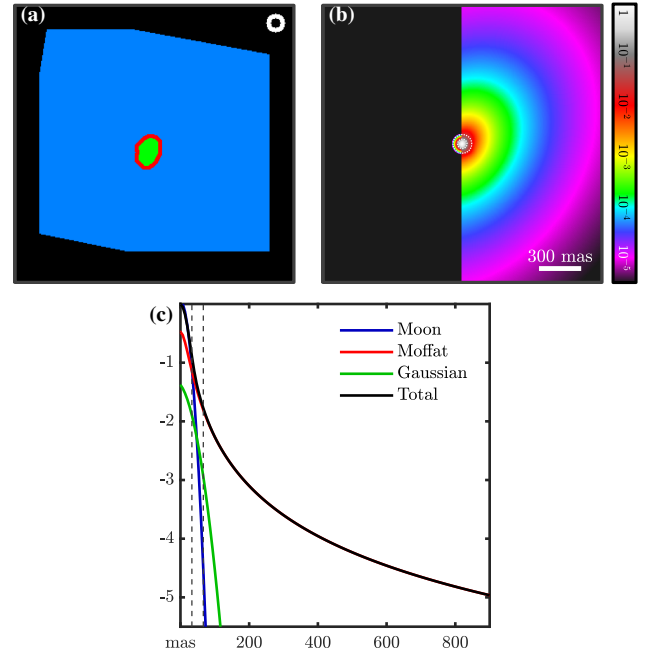
 
        \centering
        
        \newcommand{\LineRatio}{0.9}
        
        \newcommand{\PathFig}{IP_halo_}
        \newcommand{\FlagReduc}{_PIC}
        \newcommand{\FigOne}{\PathFig Mask\FlagReduc .pdf}
        \newcommand{\FigTwo}{\PathFig PSF_model\FlagReduc .pdf}
        \newcommand{\FigThree}{\PathFig fig_scale_bar.pdf}
        \newcommand{\FigFour}{\PathFig PSF_profiles\FlagReduc .pdf}
        \newcommand{\subfigColor}{white}

        \sbox1{\includegraphics{\FigOne}}               
        \sbox2{\includegraphics{\FigTwo}}               
        \sbox3{\includegraphics{\FigThree}}     
        \sbox4{\includegraphics{\FigFour}}              
        \newcommand{\ColumnWidth}[1]
                {\dimexpr \LineRatio \linewidth * \AspectRatio{#1} / (\AspectRatio{1} + \AspectRatio{2} + \AspectRatio{3}) \relax
                }
        \newcommand{\ColumnGap}{\hspace {\dimexpr \linewidth /4 - \LineRatio\linewidth /4 }}

        \begin{tabular}{
                @{\ColumnGap}
                C{\ColumnWidth{1}}
                @{\ColumnGap}
                C{\ColumnWidth{2}}
                @{}
                C{\ColumnWidth{3}}
                @{\ColumnGap}
                }
                \subfigimg[width=\linewidth,pos=ul,font=\fontfig{\subfigColor}]{$\;$(a)}{0.0}{\FigOne} &
                \subfigimg[width=\linewidth,pos=ul,font=\fontfig{\subfigColor}]{$\;$(b)}{0.0}{\FigTwo} &
                \subfigimg[width=\linewidth,pos=ul,font=\fontfig{\subfigColor}]{}{0.0}{\FigThree}
        \end{tabular}   
        
        \vspace{-1pt}
        
        \renewcommand{\LineRatio}{0.5}
        \renewcommand{\ColumnWidth}[1]
                {\dimexpr \LineRatio \linewidth * \AspectRatio{#1} / (\AspectRatio{4}) \relax
                }
        \renewcommand{\ColumnGap}{\hspace {\dimexpr \linewidth /2 - \LineRatio\linewidth /2 }}

        \begin{tabular}{
                @{\ColumnGap}
                C{\ColumnWidth{4}}
                @{\ColumnGap}
                }
                \subfigimg[width=\linewidth,pos=ul,font=\fontfig{black}]{$\!\!\!$(c)}{0.0}{\FigFour}
        \end{tabular}
        
        \caption{\label{fig:IP_halo} Fitting the PSF wings from \subfigs{fig:IP_overview}{b,c} and \subfig{fig:IP_deconv}{b}. (a)~Map of the different mask on the pixel. Green: pixels of the primary after segmentation. Red: pixels removed after the primary mask erosion. Black: pixels outside the IFS field of view. Blue: pixels on which the halo was fitted and the median filter applied. It should be noted that the annulus filter is displayed in white. (b)~Comparison of the core of the PSF~$\Vcore$ of \subfig{fig:IP_deconv}{b} (left) with the total PSF~$\Vcore + \Vwing$ fitted by accounting for the halo (right) (logarithmic scale). The dotted white lines emphasise the edges of the annulus median filter. (c)~$x$-profiles of (b) (logarithmic scale). The different components of the PSF are emphasised by different colours. Blue: PSF core~$\Vcore$ fitted on the brightest moon (normalised to its maximal value). Black: total PSF~$\Vcore + \Vwing$ (normalised to its maximal value).  Red: Moffat sub-part~$m$ of the PSF wings~$\Vwing$ (normalised to the total PSF maximal value). Green: Gaussian sub-part~$g$ of the PSF wings~$\Vwing$ (normalised to the total PSF maximal value). Dashed grey lines: edges of the annulus median filter.}
\end{figure}

Once the detailed image of the asteroid~$\Vthr$ was reconstructed, the PSF extensions~$\Vwing$ could be estimated. This is similar to deconvolving the data by the primary image to obtain the PSF wings. The model of~$\Vwing$ was fitted only on the halo. As a consequence, the pixels of the primary of~$\Vthr$ were rejected (as seen in green in \subfig{fig:IP_halo}{a}). In addition, to avoid any impact of the core of the PSF~$\Vcore$ close to the edges of the asteroid, this mask was further morphologically eroded by 5 pixels \citep{Gonzalez:20_Matlab} (in red in \subfig{fig:IP_halo}{a}). This roughly corresponds to the radius of the PSF (see \fig{fig:IP_deconv}). And, as previously done, the pixels outside the IFS field of view were also rejected. In total, only the pixels~$\Whal$ (in blue in \subfig{fig:IP_halo}{a}) were kept to fit the halo from the data~$\Vred$.

The model of the PSF wings is a summation of a 2D Gaussian function and a 2D Moffat function
\begin{align}
        \label{eq:PSF_wing}
        \Vwing\Paren{\x}
        {} = {} &
        g\Paren{\x-\x_{0}, a, \alpha, \sig}
        + m\Paren{\x-\tilde{\x}_{0}, \tilde{a}, \tilde{\alpha}, \tilde{\beta}, \tilde{\sig}}
        \\
        {} \triangleq {} &
        a \E{-0.5\Paren{
                {r_{1}^{2}}/{\sigma_{1}^{2}}
                + {r_{2}^{2}}/{\sigma_{2}^{2}}}}
        + \tilde{a} \Paren{1 + 
                {\tilde{r}_{1}^{2}}/{\sigma_{1}^{2}}
                + {\tilde{r}_{2}^{2}}/{\sigma_{2}^{2}}}^{-\tilde{\beta}}
        \,,
\end{align}
where the 2D Gaussian function $g$ and its parameters~$\Paren{a, \alpha, \sig}$ are defined as in \eq{eq:PSF_core}; $\tilde{a}$ is the amplitude of the 2D Moffat pattern; $\tilde{\alpha}$ is the orientation of the 2D Moffat pattern; $\tilde{\beta}$ is the power parameter of the 2D Moffat pattern; $\tilde{r}_{1} = \Paren{x-\tilde{x}_{0}}\cos{\tilde{\alpha}} + \Paren{y-\tilde{y}_{0}}\sin{\tilde{\alpha}}$ and $\tilde{r}_{2} = -\Paren{x-\tilde{x}_{0}}\sin{\tilde{\alpha}} + \Paren{y-\tilde{y}_{0}}\cos{\tilde{\alpha}}$ are the coordinates in the rotated frame of the Moffat pattern;  and $\tilde{\sig} = \Paren{\tilde{\sigma}_{1}, \tilde{\sigma}_{2}}$ are the extensions of the Moffat pattern along its two axes. The positions~$\x_{0}$ and~$\tilde{\x}_{0}$ were let free to account for a possible non-symmetry in the PSF wing, which is shifted compared to the PSF core.

As for the PSF core, a background can corrupt the PSF wings' estimate and it was consequently fitted within the halo model~$\Vhalo$
\begin{align}
        \Vred\Paren{\x} {}\simeq{} &
        \Vcore\Paren{\x} \conv \Vthr\Paren{\x}
        \\
        & + 
        g\Paren{\x-\x_{0}, a, \alpha, \sig}\conv \Vthr\Paren{\x}
        \\
        & +
        m\Paren{\x-\tilde{\x}_{0}, \tilde{a}, \tilde{\alpha}, \tilde{\beta}, \tilde{\sig}} \conv \Vthr\Paren{\x}
        \\
        & +
        \Vbg\Paren{\x, \V{c}}
        \\
        \label{eq:halo}
        {} \triangleq {} &
        \Vhalo\Paren{\x,\x_{0}, a, \alpha, \sig, \tilde{\x}_{0}, \tilde{a}, \tilde{\alpha}, \tilde{\beta}, \tilde{\sig}, \V{c}}
        \,,
\end{align}
where~$\simeq$ is to symbolise the uncertainties from the noise for $\Whal\Paren{\x}=1$.

Similar to the PSF core fitting in \refsec{app:IP_core}, all these parameters were estimated in an alternate algorithm. First, the 2D Moffat parameters were obtained by minimising
\begin{multline} 
        \label{eq:solve_moff_wing}
        \underset{\tilde{\x}_{0}, \tilde{a}, \tilde{\alpha}, \tilde{\beta}, \tilde{\sig}}{\text{argmin}}\;\sum_{\x \st \Whal\Paren{\x} = 1}\Big(\Vred\Paren{\x} - 
        \\
        \Vhalo\Paren{\x,\x_{0}, a, \alpha, \sig, \tilde{\x}_{0}, \tilde{a}, \tilde{\alpha}, \tilde{\beta}, \tilde{\sig}, \V{c}}\Big) ^{2}
        \,
\end{multline}
by keeping~$\Paren{\x_{0}, a, \alpha, \sig, \V{c}}$ fixed. Then, the 2D Gaussian parameters were obtained by minimising
\begin{multline} 
        \label{eq:solve_gauss_wing}
        \underset{\x_{0}, a, \alpha, \sig}{\text{argmin}}\;\sum_{\x \st \Whal\Paren{\x} = 1}\Big(\Vred\Paren{\x} - 
        \\
        \Vhalo\Paren{\x,\x_{0}, a, \alpha, \sig, \tilde{\x}_{0}, \tilde{a}, \tilde{\alpha}, \tilde{\beta}, \tilde{\sig}, \V{c}}\Big) ^{2}
        \,
\end{multline}
by keeping~$\Paren{\tilde{\x}_{0}, \tilde{a}, \tilde{\alpha}, \tilde{\beta}, \tilde{\sig}, \V{c}}$ fixed. Finally, the background parameters were estimated by solving
\begin{multline} 
        \label{eq:solve_bg_wing}
        \underset{\V{c}}{\text{argmin}}\;\sum_{\x \st \Whal\Paren{\x} = 1}\Big|\Vred\Paren{\x} - 
        \\
        \Vhalo\Paren{\x,\x_{0}, a, \alpha, \sig, \tilde{\x}_{0}, \tilde{a}, \tilde{\alpha}, \tilde{\beta}, \tilde{\sig}, \V{c}}\Big|
        \,,
\end{multline}
by keeping~$\Paren{\x_{0}, a, \alpha, \sig, \tilde{\x}_{0}, \tilde{a}, \tilde{\alpha}, \tilde{\beta}, \tilde{\sig}}$ fixed.

This algorithm is summarised in Algorithm~\ref{alg:IP_wing}. The optimisation problems seen in \eqs{eq:solve_moff_wing}{eq:solve_gauss_wing} were solved by the VMLM-B algorithm \citep{Thiebaut:02}. The optimisation problem in \eq{eq:solve_bg_wing} was solved by the simplex search method of \cite{Lagarias:98_fminsearch}.

\begin{algorithm}
\caption{\label{alg:IP_wing} Pseudo-code of the PSF wings' fitting on the halo of the primary.}
\begin{algorithmic}[1]
\small

\State $\V{c} \gets \V{0}$
        \commentalgo{Initialisation with no background}

\State $a \gets 0$
        \commentalgo{Initialisation with no Gaussian pattern}

\For{$i\text{ from } 1 \text{ to } 10$}
        \commentalgo{$Ten$ iterations of the alternate optimisation}
        
        \State $\Paren{\tilde{\x}_{0}, \tilde{a}, \tilde{\alpha}, \tilde{\beta}, \tilde{\sig}} \gets $ solving \eq{eq:solve_moff_wing}
                \commentalgo{2D Moffat pattern fitting}
                
        \State $\Paren{\x_{0}, a, \alpha, \sig} \gets $ solving \eq{eq:solve_gauss_wing}
                \commentalgo{2D Gaussian pattern fitting}
                
        \State $\V{c} \gets $ solving \eq{eq:solve_bg_wing}
                \commentalgo{Background fitting}
                        
\EndFor
\end{algorithmic}
\end{algorithm}

The results are presented in \subfigs{fig:IP_halo}{b,c}. It first appears from \subfig{fig:IP_halo}{b} that the PSF wings are elongated and not radially symmetric. The PSF orientation follows that of \Elektra's halo in \subfig{fig:IP_overview}{c}. Then, the comparison in \subfig{fig:IP_halo}{c} of the profile of the PSF core~$\Vcore$ fitted on the moon (in blue), with the profile of the final total PSF~$\Vcore + \Vwing$ in black, shows that the shape of the PSF core, below the first dashed line, is not changed by the fit of the wings. And beyond this radius, the PSF profile is dominated by its wings, mainly approximated by the Moffat profile (in red). This validates our `core + wings' approach and its fitting strategy. Finally, these curves also show that the additional 2D Gaussian pattern (in green) only plays a minor role, slightly more extended than the PSF core, meaning that it mainly impacts the halo close to the primary.

Looking at the edges of the median filter (the grey dashed lines in \subfig{fig:IP_halo}{c}), it appears that the PSF wing intensity is a few percent of its maximal value in the filter domain. This is consistent with the residual structures inside the median filter mask on \subfig{fig:IP_deconv}{c}, beyond the PSF core. In case one wants to obtain the absolute photometry, a constant correction factor must be applied that can be fitted on the PSF model. This is nonetheless beyond the scope of this Letter and does impact our photometry study based on photometric ratios.

\subsection{Median filter}
\label{app:IP_median}

Once the total PSF was estimated as described in \refapp{app:IP_halo}, the model of the halo~$\Vhalo$ defined in \eq{eq:halo} was removed from the reduced data
\begin{equation}
        \Vres\Paren{\x} = \Vred\Paren{\x} - \Vhalo\Paren{\x,\x_{0}, a, \alpha, \sig, \tilde{\x}_{0}, \tilde{a}, \tilde{\alpha}, \tilde{\beta}, \tilde{\sig}, \V{c}}
        \,.
\end{equation}
These residuals are shown in \subfig{fig:IP_overview}{e}. Most of the halo was successfully removed, as expected, close to the primary edges and the moons are better visible. Nonetheless some spatially extended structures remain that can be attributed to the instrument background or the sky background.

To further clean these artefacts, a median filter was applied on this residual map. To prevent any self-subtraction, the shape of this filter is an annulus whose inner (resp. outer) diameter is once (resp. twice) the PSF core size. The idea is to correct for local background features, hence the disc shape, while preventing self-subtraction of the moon due to the core of the PSF, hence the annulus. The shape of this annulus is given in \subfig{fig:IP_deconv}{c} and \subfigs{fig:IP_halo}{a,b}. It must be compared with the PSF core size: its inner diameter is~$5$ pixels, corresponding to the size of the PSF core, and its outer diameter is twice this value, that is to say $10$ pixels.

This median filter is applied only on the pixels given by~$\Whal$, in blue in \subfig{fig:IP_halo}{a}, as described in Algorithm~\ref{alg:IP_median}.

\begin{algorithm}
\caption{\label{alg:IP_median} Pseudo-code of the median filtering with an annulus mask.}
\begin{algorithmic}[1]
\small

\State $\Vfilt \gets \V{0}$
        \commentalgo{Initialisation with zeros}
        
\State $r^{\Tag{in}} \gets 5$
        \commentalgo{Annulus mask's inner radius}
        
\State $r^{\Tag{out}} \gets 10$
        \commentalgo{Annulus mask's outer radius}

\For{$\x \st \Whal\Paren{\x}=1$}
        \commentalgo{Filtering only the pixels in~$\Whal$}
        
        \State $\medval \gets \text{med} \Brack{
                \Vres\Paren{\tilde{\x} \st
                \begin{cases}
                \Whal\Paren{\tilde{\x}}=1
                \\
                r^{\Tag{in}} \leq \Norm{\tilde{\x}-\x} \leq r^{\Tag{out}}
                \end{cases}}
                }$
                \commentalgo{Median value}
        
        \State $\Vfilt\Paren{\x} \gets \Vres\Paren{\x} - \medval$
                \commentalgo{Removing the median filter}
                                        
\EndFor
\end{algorithmic}
\end{algorithm}

The result of this median filtering, $\Vfilt$, is given in \subfig{fig:IP_overview}{f}. Most of the background structures are efficiently removed. The only remaining artefacts have a size similar to the PSF core. The three moons are now clearly visible in a single acquisition frame.

\FloatBarrier

\section{Comparison with the SPHERE/DC reduction pipeline}
\label{app:DC}


For comparison purposes, we present in this section the results obtained using the reduction pipeline of the SPHERE Data Centre \citep[SPHERE/DC,][]{Delorme:17_SPHERE_pipeline}. The same image processing steps were applied on the hypercubes to estimate and remove the halo.

\begin{figure}[ht!]
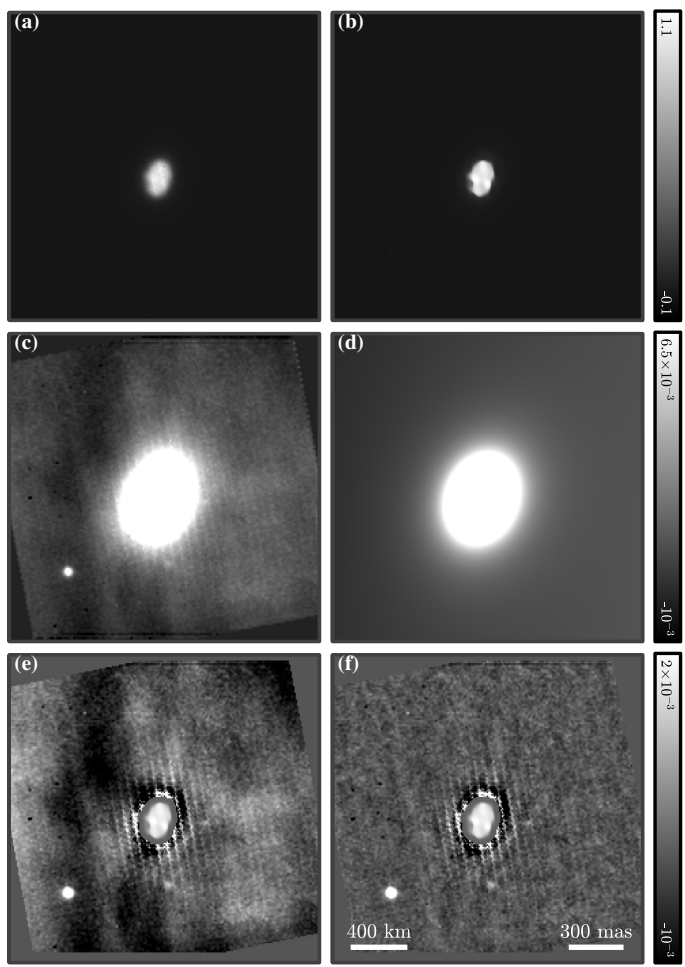
 
        \centering
        
        \newcommand{\LineRatio}{1}
        
        \newcommand{\PathFig}{IP_overview_}
        \newcommand{\FlagReduc}{_DC}
        \newcommand{\FigOne}{\PathFig Step1_raw_data_obj\FlagReduc .pdf}
        \newcommand{\FigTwo}{\PathFig Step2_deconvolution_obj\FlagReduc .pdf}
        \newcommand{\FigThree}{\PathFig Step1_raw_data_obj_bar.pdf}
        \newcommand{\subfigColor}{white}        
        
        \sbox1{\includegraphics{\FigOne}}               
        \sbox2{\includegraphics{\FigTwo}}               
        \sbox3{\includegraphics{\FigThree}}     
        \newcommand{\ColumnWidth}[1]
                {\dimexpr \LineRatio \linewidth * \AspectRatio{#1} / (\AspectRatio{1} + \AspectRatio{2} + \AspectRatio{3}) \relax
                }
        \newcommand{\ColumnGap}{\hspace {\dimexpr \linewidth /4 - \LineRatio\linewidth /4 }}

        \begin{tabular}{
                @{\ColumnGap}
                @{\ColumnGap}
                C{\ColumnWidth{1}}
                @{\ColumnGap}
                C{\ColumnWidth{2}}
                @{\ColumnGap}
                C{\ColumnWidth{3}}
                @{\ColumnGap}
                }
                \subfigimg[width=\linewidth,pos=ul,font=\fontfig{\subfigColor}]{$\;$(a)}{0.0}{\FigOne} &
                \subfigimg[width=\linewidth,pos=ul,font=\fontfig{\subfigColor}]{$\;$(b)}{0.0}{\FigTwo} &
                \subfigimg[width=\linewidth,pos=ul,font=\fontfig{\subfigColor}]{}{0.0}{\FigThree}
        \end{tabular}
        
        \vspace{-5pt}
        
        \renewcommand{\FigOne}{\PathFig Step1_raw_data_halo\FlagReduc .pdf}
        \renewcommand{\FigTwo}{\PathFig Step3_fitting_halo_mod\FlagReduc .pdf}
        \renewcommand{\FigThree}{\PathFig Step1_raw_data_halo_bar.pdf}
        
        \sbox1{\includegraphics{\FigOne}}               
        \sbox2{\includegraphics{\FigTwo}}               
        \sbox3{\includegraphics{\FigThree}}     

        \begin{tabular}{
                @{\ColumnGap}
                C{\ColumnWidth{1}}
                @{\ColumnGap}
                C{\ColumnWidth{2}}
                @{\ColumnGap}
                C{\ColumnWidth{3}}
                @{\ColumnGap}
                }
                \subfigimg[width=\linewidth,pos=ul,font=\fontfig{\subfigColor}]{$\;$(c)}{0.0}{\FigOne} &
                \subfigimg[width=\linewidth,pos=ul,font=\fontfig{\subfigColor}]{$\;$(d)}{0.0}{\FigTwo} &
                \subfigimg[width=\linewidth,pos=ul,font=\fontfig{\subfigColor}]{}{0.0}{\FigThree}
        \end{tabular}
        
        \vspace{-5pt}
        
        \renewcommand{\FigOne}{\PathFig Step3_fitting_halo_res\FlagReduc .pdf}
        \renewcommand{\FigTwo}{\PathFig Step4_annulus_median\FlagReduc .pdf}
        \renewcommand{\FigThree}{\PathFig Step3_fitting_halo_res_bar.pdf}
        
        \sbox1{\includegraphics{\FigOne}}               
        \sbox2{\includegraphics{\FigTwo}}               
        \sbox3{\includegraphics{\FigThree}}     

        \begin{tabular}{
                @{\ColumnGap}
                C{\ColumnWidth{1}}
                @{\ColumnGap}
                C{\ColumnWidth{2}}
                @{\ColumnGap}
                C{\ColumnWidth{3}}
                @{\ColumnGap}
                }
                \subfigimg[width=\linewidth,pos=ul,font=\fontfig{\subfigColor}]{$\;$(e)}{0.0}{\FigOne} &
                \subfigimg[width=\linewidth,pos=ul,font=\fontfig{\subfigColor}]{$\;$(f)}{0.0}{\FigTwo} &
                \subfigimg[width=\linewidth,pos=ul,font=\fontfig{\subfigColor}]{}{0.0}{\FigThree}
        \end{tabular}

        \caption{\label{fig:IP_overview_DC} Overview of the image processing steps, applied on the $24^{\text{th}}$ acquisition of 9 Dec 2014 after its reduction by the SPHERE/DC pipeline \citep{Delorme:17_SPHERE_pipeline}. See caption of \fig{fig:IP_overview}.}
\end{figure}

The reduced projection in \subfig{fig:IP_overview_DC}{a} presents vertical artefacts that corrupt the image deconvolution. They are interpreted as a misalignment between the calibrations and the science images \citep{Berdeu:20_PIC}, a parameter that is estimated, and thus corrected, by the PIC reduction pipeline.

Comparing \subfig{fig:IP_overview_DC}{b} with \subfig{fig:IP_overview}{b}, it is clear that these artefacts induce the reconstructed image of the asteroid surface to be less smooth and homogeneous than with the PIC reduction pipeline. This can impact the halo fitting accuracy. But above all, this can bias the integrated flux from the primary and its centroiding, degrading the quality of the astrophotometry fit.

These artefacts are also clearly seen in the saturated view given in \subfig{fig:IP_overview_DC}{c}. They cannot be removed by the halo fitting (see \subfig{fig:IP_overview_DC}{e}), nor by the median filtering (see \subfig{fig:IP_overview_DC}{f}). As a consequence, the faint third moon can be mistaken with these artefacts. On the contrary, this moon is obvious in the data processed after the reduction by PIC (in \subfig{fig:IP_overview}{f}).

Comparing the time-lapses of the data reduction and processing of {\textcolor{blue}{Visualisations 1}} and {\textcolor{blue}{5}} leads to the same conclusions. The PIC pipeline provides a smoother and less noisy background than the SPHERE/DC pipeline. The third moon motion is also barely visible behind the parallel artefacts.

Comparing the residuals of the PSF core fitting in \subfig{fig:IP_deconv}{c} and \subfig{fig:IP_deconv_DC}{c}, it clearly appears that the reduction with the PIC pipeline is less noisy. Some features in the PSF wing are even visible in \subfig{fig:IP_deconv}{c}, while being completely erased in the noise with the SPHERE/DC reduction in \subfig{fig:IP_deconv_DC}{c}. This further confirms that adequate regularisations and priors in the IFS raw data reduction can push forward the instrument contrast and detection limits.

\begin{figure}[ht!] 
        \centering
        
        \newcommand{\LineRatio}{1}
        
        \newcommand{\PathFig}{IP_deconv_}
        \newcommand{\FlagReduc}{_DC}
        \newcommand{\FigOne}{\PathFig PSF\FlagReduc .pdf}
        \newcommand{\FigTwo}{\PathFig PSF_model\FlagReduc .pdf}
        \newcommand{\FigThree}{\PathFig PSF_bar.pdf}
        \newcommand{\FigFour}{\PathFig PSF_residuals\FlagReduc .pdf}
        \newcommand{\FigFive}{\PathFig PSF_residuals_bar.pdf}
        \newcommand{\subfigColor}{white}        
        
        \sbox1{\includegraphics{\FigOne}}               
        \sbox2{\includegraphics{\FigTwo}}               
        \sbox3{\includegraphics{\FigThree}}     
        \sbox4{\includegraphics{\FigFour}}      
        \sbox5{\includegraphics{\FigFive}}      
        \newcommand{\ColumnWidth}[1]
                {\dimexpr \LineRatio \linewidth * \AspectRatio{#1} / (\AspectRatio{1} + \AspectRatio{2} + \AspectRatio{3} + \AspectRatio{4} + \AspectRatio{5}) \relax
                }
        \newcommand{\ColumnGap}{\hspace {\dimexpr \linewidth /6 - \LineRatio\linewidth /6 }}

        \begin{tabular}{
                @{\ColumnGap}
                C{\ColumnWidth{1}}
                @{\ColumnGap}
                C{\ColumnWidth{2}}
                @{\ColumnGap}
                C{\ColumnWidth{3}}
                @{\ColumnGap}
                C{\ColumnWidth{4}}
                @{\ColumnGap}
                C{\ColumnWidth{5}}
                @{\ColumnGap}
                }
                \subfigimg[width=\linewidth,pos=ul,font=\fontfig{\subfigColor}]{$\,$(a)}{0.0}{\FigOne} &
                \subfigimg[width=\linewidth,pos=ul,font=\fontfig{\subfigColor}]{$\,$(b)}{0.0}{\FigTwo} &
                \subfigimg[width=\linewidth,pos=ul,font=\fontfig{\subfigColor}]{}{0.0}{\FigThree} &
                \subfigimg[width=\linewidth,pos=ul,font=\fontfig{\subfigColor}]{$\,$(c)}{0.0}{\FigFour} &
                \subfigimg[width=\linewidth,pos=ul,font=\fontfig{\subfigColor}]{}{0.0}{\FigFive}
        \end{tabular}   
        
        \caption{\label{fig:IP_deconv_DC} Fitting the core of the PSF on the brightest moon from \subfig{fig:IP_overview_DC}{c}. See caption of \fig{fig:IP_deconv}.}
\end{figure}

\figfull{fig:IP_halo_DC} presents the fit of the PSF wings. It shows that it is consistent with the fit obtained for the data reduced with PIC in \fig{fig:IP_halo}. This supports the fact that our parametric method, which depends on a limited number of parameters to describe the PSF, is robust to local artefacts.

\begin{figure}[ht!] 
        \centering
        
        \newcommand{\LineRatio}{1}
        
        \newcommand{\PathFig}{IP_halo_}
        \newcommand{\FlagReduc}{_DC}
        \newcommand{\FigOne}{\PathFig Mask\FlagReduc .pdf}
        \newcommand{\FigTwo}{\PathFig PSF_model\FlagReduc .pdf}
        \newcommand{\FigThree}{\PathFig fig_scale_bar.pdf}
        \newcommand{\FigFour}{\PathFig PSF_profiles\FlagReduc .pdf}
        \newcommand{\subfigColor}{white}        
        
        \sbox1{\includegraphics{\FigOne}}               
        \sbox2{\includegraphics{\FigTwo}}               
        \sbox3{\includegraphics{\FigThree}}     
        \sbox4{\includegraphics{\FigFour}}              
        \newcommand{\ColumnWidth}[1]
                {\dimexpr \LineRatio \linewidth * \AspectRatio{#1} / (\AspectRatio{1} + \AspectRatio{2} + \AspectRatio{3}) \relax
                }
        \newcommand{\ColumnGap}{\hspace {\dimexpr \linewidth /4 - \LineRatio\linewidth /4 }}

        \begin{tabular}{
                @{\ColumnGap}
                C{\ColumnWidth{1}}
                @{\ColumnGap}
                C{\ColumnWidth{2}}
                @{}
                C{\ColumnWidth{3}}
                @{\ColumnGap}
                }
                \subfigimg[width=\linewidth,pos=ul,font=\fontfig{\subfigColor}]{$\;$(a)}{0.0}{\FigOne} &
                \subfigimg[width=\linewidth,pos=ul,font=\fontfig{\subfigColor}]{$\;$(b)}{0.0}{\FigTwo} &
                \subfigimg[width=\linewidth,pos=ul,font=\fontfig{\subfigColor}]{}{0.0}{\FigThree}
        \end{tabular}   
        
        \vspace{-1pt}
        
        \renewcommand{\LineRatio}{0.5}
        \renewcommand{\ColumnWidth}[1]
                {\dimexpr \LineRatio \linewidth * \AspectRatio{#1} / (\AspectRatio{4}) \relax
                }
        \renewcommand{\ColumnGap}{\hspace {\dimexpr \linewidth /2 - \LineRatio\linewidth /2 }}

        \begin{tabular}{
                @{\ColumnGap}
                C{\ColumnWidth{4}}
                @{\ColumnGap}
                }
                \subfigimg[width=\linewidth,pos=ul,font=\fontfig{black}]{$\!\!\!$(c)}{0.0}{\FigFour}
        \end{tabular}
        
        \caption{\label{fig:IP_halo_DC} Fitting the PSF wings from \subfigs{fig:IP_overview_DC}{b,c} and \subfig{fig:IP_deconv_DC}{b}. See caption of \fig{fig:IP_halo}.}
\end{figure}

\begin{figure}[ht!]
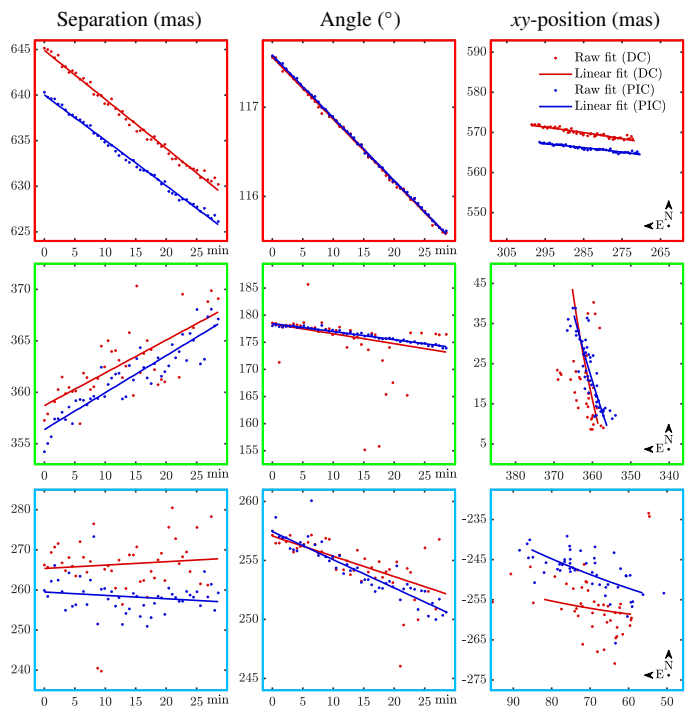
 
        \centering
        
        \newcommand{\LineRatio}{1}
        \newcommand{\PathFig}{DCvsPIC_}
        
        \newcommand{\FlagMoon}{m1}
        \newcommand{\FigOne}{\PathFig Elektra_2014-12-09_sep_\FlagMoon .pdf}
        \newcommand{\FigTwo}{\PathFig Elektra_2014-12-09_ang_\FlagMoon .pdf}
        \newcommand{\FigThree}{\PathFig Elektra_2014-12-09_xy_\FlagMoon .pdf}
        \newcommand{\subfigColor}{white}        
        
        \sbox1{\includegraphics{\FigOne}}               
        \sbox2{\includegraphics{\FigTwo}}               
        \sbox3{\includegraphics{\FigThree}}     
        \newcommand{\ColumnWidth}[1]
                {\dimexpr \LineRatio \linewidth * \AspectRatio{#1} / (\AspectRatio{1} + \AspectRatio{2} + \AspectRatio{3}) \relax
                }
        \newcommand{\ColumnGap}{\hspace {\dimexpr \linewidth /4 - \LineRatio\linewidth /4 }}

        \begin{tabular}{
            @{\ColumnGap}
                C{\ColumnWidth{1}}
                @{\ColumnGap}
                C{\ColumnWidth{2}}
                @{\ColumnGap}
                C{\ColumnWidth{3}}
                @{\ColumnGap}
                }
                \tiny $\;\;\;$ Separation (\mas{}) &
                \tiny $\;\;\;\;$ Angle (\degr) &
                \tiny $\;\;\;$ $xy$-position (\mas{})
                \\
                \subfigimg[width=\linewidth,pos=ul,font=\fontfig{\subfigColor}]{}{0.0}{\FigOne} &
                \subfigimg[width=\linewidth,pos=ul,font=\fontfig{\subfigColor}]{}{0.0}{\FigTwo} &
                \subfigimg[width=\linewidth,pos=ul,font=\fontfig{\subfigColor}]{}{0.0}{\FigThree}
        \end{tabular}
        
        \vspace{-5pt}
        
        \renewcommand{\FlagMoon}{m2}
        \renewcommand{\FigOne}{\PathFig Elektra_2014-12-09_sep_\FlagMoon .pdf}
        \renewcommand{\FigTwo}{\PathFig Elektra_2014-12-09_ang_\FlagMoon .pdf}
        \renewcommand{\FigThree}{\PathFig Elektra_2014-12-09_xy_\FlagMoon .pdf}
        \renewcommand{\subfigColor}{white}      
        
        \sbox1{\includegraphics{\FigOne}}               
        \sbox2{\includegraphics{\FigTwo}}               
        \sbox3{\includegraphics{\FigThree}}     
        \renewcommand{\ColumnWidth}[1]
                {\dimexpr \LineRatio \linewidth * \AspectRatio{#1} / (\AspectRatio{1} + \AspectRatio{2} + \AspectRatio{3}) \relax
                }
        \renewcommand{\ColumnGap}{\hspace {\dimexpr \linewidth /4 - \LineRatio\linewidth /4 }}
        
        \begin{tabular}{
                @{\ColumnGap}
                C{\ColumnWidth{1}}
                @{\ColumnGap}
                C{\ColumnWidth{2}}
                @{\ColumnGap}
                C{\ColumnWidth{3}}
                @{\ColumnGap}
                }
                \subfigimg[width=\linewidth,pos=ul,font=\fontfig{\subfigColor}]{}{0.0}{\FigOne} &
                \subfigimg[width=\linewidth,pos=ul,font=\fontfig{\subfigColor}]{}{0.0}{\FigTwo} &
                \subfigimg[width=\linewidth,pos=ul,font=\fontfig{\subfigColor}]{}{0.0}{\FigThree}
        \end{tabular}
        \vspace{-5pt}
        
        \renewcommand{\FlagMoon}{m3}
        \renewcommand{\FigOne}{\PathFig Elektra_2014-12-09_sep_\FlagMoon .pdf}
        \renewcommand{\FigTwo}{\PathFig Elektra_2014-12-09_ang_\FlagMoon .pdf}
        \renewcommand{\FigThree}{\PathFig Elektra_2014-12-09_xy_\FlagMoon .pdf}
        \renewcommand{\subfigColor}{white}      
        
        \sbox1{\includegraphics{\FigOne}}               
        \sbox2{\includegraphics{\FigTwo}}               
        \sbox3{\includegraphics{\FigThree}}     
        \renewcommand{\ColumnWidth}[1]
                {\dimexpr \LineRatio \linewidth * \AspectRatio{#1} / (\AspectRatio{1} + \AspectRatio{2} + \AspectRatio{3}) \relax
                }
        \renewcommand{\ColumnGap}{\hspace {\dimexpr \linewidth /4 - \LineRatio\linewidth /4 }}
        
        \begin{tabular}{
                @{\ColumnGap}
                C{\ColumnWidth{1}}
                @{\ColumnGap}
                C{\ColumnWidth{2}}
                @{\ColumnGap}
                C{\ColumnWidth{3}}
                @{\ColumnGap}
                }
                \subfigimg[width=\linewidth,pos=ul,font=\fontfig{\subfigColor}]{}{0.0}{\FigOne} &
                \subfigimg[width=\linewidth,pos=ul,font=\fontfig{\subfigColor}]{}{0.0}{\FigTwo} &
                \subfigimg[width=\linewidth,pos=ul,font=\fontfig{\subfigColor}]{}{0.0}{\FigThree}
        \end{tabular}
        
        \caption{\label{fig:DCvsPIC} Comparison of the moon position fitting for the SPHERE/DC (dark red) and PIC pipelines (dark blue) for the three moons: \Sone (first line, red frame), \Stwo (second line, green frame), and \Sthree (third line, blue frame). The positions are given in polar coordinates,  separation (first column), and angle (second column) as well as Cartesian coordinates ($xy$-positions, third column). The dots are the position individually fitted in each acquisition. The curves are the temporal linear fits performed on the polar coordinates.}
\end{figure}

Finally, \fig{fig:DCvsPIC} summarises the performances of the moon fitting strategy applied on the processed data after their reduction with the two pipelines. It first appears that the dispersion of the moon positions fitted on each frame (dots) is higher with the SPHERE/DC pipeline (dark red) than with PIC (dark blue). This is particularly visible on the angle (middle column): with the PIC pipeline, the temporal evolution follows a monotonous trend whereas the SPHERE/DC pipeline produces a lot more outliers. This can also be seen in the $xy$-position plane where the dispersion around the polar linear fit (plain curve) is reduced with PIC. Then, looking at the separation and $xy$-position evolution of \Sone (red) and \Stwo (green), a constant bias can be noticed between the two pipelines. These biases correspond to a shift in the linear fits of $\Paren{\delta x, \delta y} \simeq \Paren{1.8,-3.9}\,\mas{}$ for \Sone and $\Paren{\delta x, \delta y} \simeq \Paren{1.5,-3.6}\,\mas{}$ for \Stwo. The similarity of these values allows one to argue in favour of a bias in the primary centre estimation rather than a plate scale error, which would produce a shift proportional with the distance to the primary. As mentioned earlier, the artefacts of the SPHERE/DC pipeline are present in the deconvolved image of \Elektra and can bias the estimate of its photocentre. Finally, the offset between the two pipelines on \Sthree (blue), $\Paren{\delta x, \delta y} \simeq \Paren{-0.3,8.7}\,\mas{}$, can be attributed to the stripe-shaped artefacts in the SPHERE/DC reduction that bias the estimation of the moon position. This can also be seen in {\textcolor{blue}{Visualisation 5}} where the motion of \Sthree seems to follow one from this artefact stripe.

\FloatBarrier

\section{Discussion on the orbital fit}

In this appendix, we discuss the twofold degeneracy in the orbital pole fit and we list the orbital elements of \Sone and \Stwo as well as the spin alignment of the different moons.

\subsection{Degeneracy of the orbital pole}
\label{app:Orbit_fitting_pole}

With datasets spanning from 9 Dec  to 31 Dec 2014, it could be expected that the change in the viewing angle of the system, of only $3.6\degr$, is insufficient to lift the twofold degeneracy in the orbital pole. This degeneracy originates from the projection of a 3D ellipse onto the 2D plane tangent to the observation direction.

Concerning the degeneracy on \Sone, we recall here that the system mass and the orbital parameters of \Sone were initialised on archival data spanning from 2003 to 2019, as detailed in \tab{tab:Sone_data}. In these datasets, the system was viewed from many angles and the orbital pole of \Sone was unequivocally determined. This solution was thus kept to fit the orbits of \Stwo and \Sthree in the 2014 datasets.

\begin{table}[ht!] 
    \begin{center}
        \caption{\label{tab:Sone_data} List of the archival data used for the orbital fit initialisation.}
            \tiny
            \begin{tabular}{|c|c|c|c|}
                \hline
                Year $\Paren{n_{\Tag{obs}}}$ & Instrument & PI & Programme \\
                \hline
                2003 (1) & Keck II / NIRC2 & Merline B. & N22N2 \\
                \hline
                2003 (1) & Keck II / NIRC2 & de Pater I. & U37N2 \\
                \hline
                2004 (3) & VLT / NACO & Marchis F. & 072.C-0016(A) \\
                \hline
                2004 (2) & VLT / NACO & Merline B. & 072.C-0753(A) \\
                \hline
                2004 (2) & Gem-N / NIRI & Merline B. & GN-2004B-C-5 \\
                \hline
                2005 (2) & Keck II / NIRC2 & de Pater I. & U58N2 \\
                \hline
                2006 (1) & Gem-N / NIRI & Berthier J. & GN-2006A-Q-75 \\
                \hline
                2006 (2) & VLT / NACO & Marchis F. & 077.C-0422(A) \\
                \hline
                2008 (1) & Keck II / NIRC2 & \makecell{Conrad A. \& \\ Merline B. \\ (Armandroff T.)} & K208N2L \\
                \hline
                2012 (1) & VLT / NACO & Marchis F. & 089.C-0944(B) \\
                \hline
                2016 (1) & VLT / NACO & Carry B. & 095.C-0618(B) \\
                \hline
                2019 (6) & VLT / ZIMPOL & Vernazza P. & 199.C-0074(E) \\
                \hline
            \end{tabular}
            \tablefoot{We note that $n_{\Tag{obs}}$ is the number of observations used in each observation programme directed by the principal investigator (PI).}
        \end{center}
\end{table}

In the main text in \refsec{sec:Orb_fitting}, we provide the orbital elements of \Sthree fitted by Genoid. They correspond to the best solution found by the genetic-based algorithm, looking at all the possible combinations of the different parameters. This means that one orbital pole solution was chosen by the algorithm for \Stwo and \Sthree.

In the following, we forced Genoid to fit the best solution on each orbital pole for \Stwo and \Sthree to quantify their orbital pole degeneracy. We checked if the four possible symmetries are equivalent or if the one given in this Letter indeed prevails. The results are given in \tab{tab:Pole_fitting}.  In the following, for a given moon S, S denotes the solution presented in this Letter, and \sym{S} is for its pole orbit symmetry.

\begin{table}[ht!] 
    \newcommand{\dSpin}{$\delta_{\uparrow}$\xspace}
    \begin{center}
        \caption{\label{tab:Pole_fitting} Orbital pole fit for the different symmetry on \Stwo and \Sthree.}
            \begin{tabular}{|c|c|c|c|}
                \cline{2-4}
                \multicolumn{1}{c|}{} & \dSpin (\degr) & RMS ($\mas{}$) & $\Chired$ \\
                \hline
                \Sone{} / \Stwo{} / \Sthree{} & $6 \degr / 4 \degr / 38 \degr $ & $6.77$ & $1.34$ \\
                \hline
                \Sone{} / \sym{\Stwo{}} / \Sthree{} & $ 5 \degr / 116 \degr / 39 \degr $ & $7.02$ & $1.53$ \\
                \hline
                \Sone{} / \Stwo{} / \sym{\Sthree{}} & $ 6 \degr / 2 \degr / 70 \degr $ & $8.48$ & $1.98$ \\
                \hline
                \Sone{} / \sym{\Stwo{}} / \sym{\Sthree{}} & $ 5 \degr / 118 \degr / 77 \degr $ & $9.03$ & $2.25$ \\
                \hline
            \end{tabular}
            \tablefoot{We note that \dSpin is the relative orientation of the moon pole orbit and \Elektra's spin axis. For a given moon S, S denotes the solution presented in this Letter, and \sym{S} is for its pole orbit symmetry. Furthermore, $\Chired$ is the reduced $\chi^{2}$ statistic of the fit.}
            
            
              
            
            
            
        \end{center}
\end{table}

It appears that the quality of the astrometry obtained after the robust linear fit helps Genoid to favour one solution out of the three. Concerning \Sthree (whatever the symmetry on \Stwo{}), the reduced $\chi^{2}$ statistic, $\Chired$, on \sym{\Sthree{}} is $\geq\SI{45}{\percent}$ worse than the one on \Sthree{}. Genoid thus strongly pulls towards the \Sthree symmetry, rejecting \sym{\Sthree{}}. For \Stwo (whatever the symmetry on \Sthree{}), the symmetry breaking is less obvious, the $\Chired$ being only slightly better for the \Stwo symmetry than for \sym{\Stwo{}} by $\sim\SI{15}{\percent}$. Similar conclusions were obtained looking at the RMS error.

Beyond the reduced $\Chired$ optimal value, other physics-based arguments support the \Stwo{}/\Sthree{} symmetry. Enforcing the \sym{\Stwo{}} symmetry, whatever the symmetry on \Sthree{}, leads to an orientation of \Stwo's pole orbit relative to \Elektra's spin axis of about $\sim 117\degr$, that is to say a retrograde and polar orbit that is highly improbable. Then, whatever the symmetry on \Stwo{}, the \Sthree{} symmetry solution chosen by Genoid is very inclined ($\sim 38.5 \degr$) compared to \Elektra's spin axis, but still twice less than its symmetry solution \sym{\Sthree{}}  ($\sim 73.5 \degr$), which is once again an unfavourable polar orbit situation. Finally, with this \Stwo{}/\Sthree{} solution, the three moons revolve in the same sense as \Elektra's spin, which is a reassuring feature. This solution is the one presented in this Letter.

\begin{table}[!ht]
\begin{center}
  \caption{Orbital elements of \Sone and \Stwo, 
    expressed in EQJ2000.}
    \label{tab:Orb_fitting_S1S2}
    
    \Sone{} -- \SoneFull
    \vspace{5pt}
    
  \begin{tabular}{l ll ll}
    \hline\hline
    \noalign{\smallskip}
    \multicolumn{2}{c}{Observing data set} \\
    \noalign{\smallskip}
    Number of observations & 150   \\
    Time span (days)       & 22 &  \\
    RMS (mas)              & 3.53  \\
    \hline
    \noalign{\smallskip}
    \multicolumn{2}{c}{Orbital elements EQJ2000} \\
    \noalign{\smallskip}
    $P$ (day)        &   5.287      & $\pm$  0.004    \\
    $a$ (km)         &   1353       & $\pm$  17       \\
    $e$              &   0.09       & $\pm$  0.02     \\
    $i$ (\degr)      &   161        & $\pm$  1        \\
    $\Omega$ (\degr) &   179        & $\pm$  3        \\
    $\omega$ (\degr) &   179        & $\pm$  7        \\
    $t_\Tag{p}$ (JD)     & 2456990.85   & $\pm$  0.08     \\
    $\paren{\alphap,\deltap}$ (\degr) & $\Paren{89, -71}$ & $\Paren{\pm 3, \pm 1}$       \\
    $\paren{\lambdap,\betap}$ (\degr) & $\Paren{277, -87}$ & $\Paren{\pm 20, \pm 2}$       \\
    \hline
  \end{tabular}
  
  \vspace{0.5cm}
  
    \Stwo{} -- \StwoFull
    \vspace{5pt}
  
  \begin{tabular}{l ll ll}
    \hline\hline
    \noalign{\smallskip}
    \multicolumn{2}{c}{Observing data set} \\
    \noalign{\smallskip}
    Number of observations & 120   \\
    Time span (days)       & 22 &  \\
    RMS (mas)              & 2.39  \\
    \hline
    \noalign{\smallskip}
    \multicolumn{2}{c}{Orbital elements EQJ2000} \\
    \noalign{\smallskip}
    $P$ (day)        &    1.192     & $\pm$ 0.002   \\
    $a$ (km)         &    501       & $\pm$ 7       \\
    $e$              &    0.03      & $\pm$ 0.03    \\
    $i$ (\degr)      &    156       & $\pm$  7      \\
    $\Omega$ (\degr) &    187       & $\pm$  10     \\
    $\omega$ (\degr) &    235       & $\pm$  18     \\
    $t_\Tag{p}$ (JD)     & 2456990.53   & $\pm$ 0.06    \\
    $\paren{\alphap,\deltap}$ (\degr) & $\Paren{97, -66}$ & $\Paren{\pm 10, \pm 7}$       \\
    $\paren{\lambdap,\betap}$ (\degr) & $\Paren{165, -87}$ & $\Paren{\pm 113, \pm 4}$       \\
    \hline
  \end{tabular}
    \tablefoot{Orbital period $P$, semi-major axis $a$, eccentricity $e$, inclination $i$, longitude of the ascending node $\Omega$, argument of pericentre $\omega$, time of pericentre $t_\Tag{p}$, ecliptic J2000 coordinates of the orbital pole $\paren{\lambdap,\betap}$, and equatorial J2000 coordinates of the orbital pole $\paren{\alphap,\deltap}.$ The number of observations and RMS between the predicted and observed positions are also provided. Uncertainties are given at 1-$\sigma$.}
\end{center}
\end{table}

\subsection{Orbital elements of \Sone and \Stwo}
\label{app:Orbit_fitting_S1S2}

The parameters of \Stwo were already fitted on the same 2014 datasets by \cite{Yang:16_Elektra_Minerva}. As the fit presented in this Letter gives a lower RMS of \mas{2.4} on the moon orbit model, we present this refined set of dynamical parameters for the orbital elements of \Stwo in \tab{tab:Orb_fitting_S1S2}.

On the other side, the elements of \Sone in \tab{tab:Orb_fitting_S1S2} must be handled with care. Indeed, they correspond to the fit performed in this Letter, only on the 2014 datasets, despite being initialised on data spanning from 2003 to 2019 as discussed in \refapp{app:Orbit_fitting_pole}. Using these additional observations, spanning over a period of 16 years, to perform a global fit would provide more precise parameters. However, this is beyond the scope of this Letter.

\subsection{Spin alignment}
\label{app:Orbit_fitting_spin}

\tab{tab:Spin_fitting} presents the spin alignment of all components of the \Elektra system in the global fit obtained by Genoid for the symmetry presented in this Letter $\Paren{\Stwo{} / \Sthree{}}$.

\begin{table}[!ht]
\newcommand{\spax}{$S.A.$\xspace}
\newcommand{\porb}{$P.O.$\xspace}

\begin{center}
  \caption{\label{tab:Spin_fitting}Spin alignment of all components of \Elektra.}

  \begin{tabular}{l ll ll}
    \hline\hline
    \noalign{\smallskip}
    \multicolumn{2}{c}{Spin alignment} \\
    \noalign{\smallskip}
    \spax Elektra vs \porb \Sone & 6 \degr & $\pm$ 1 \degr     \\
    \spax Elektra vs \porb \Stwo & 4 \degr & $\pm$ 5 \degr    \\
    \spax Elektra vs \porb \Sthree & 38 \degr & $\pm$ 19 \degr \\
    \porb \Sone vs \porb \Stwo & 5 \degr & $\pm$ 5 \degr      \\
    \porb \Sone vs \porb \Sthree & 40 \degr & $\pm$ 21 \degr   \\
    \porb \Stwo vs \porb \Sthree & 42 \degr & $\pm$ 20 \degr   \\
    \hline
  \end{tabular}
  \tablefoot{Given at initial condition at Julian date: 2456990. For \Elektra, we use ($\lambda,\,\beta$) = ($64 \degr,\,-88 \degr$) as the ecliptic J2000 longitude and latitude, respectively, of the spin axis \spax available on the Database of Asteroid Models from Inversion Techniques \citep[DAMIT,][]{Durech:10_Damit}, and pole orbit \porb of a satellite.Uncertainties are given at 1-$\sigma$.}
\end{center}
\end{table}

\FloatBarrier

\section{Hints of detections in other datasets}
\label{app:Extra_dates}

IRDIS/IFS and ZIMPOL are not sensitive to the same wavelength range. This makes them complementary instruments to study asteroid systems.

\textbf{IRDIS/IFS --} On one side, as described in this Letter, the more extended spatial structure of the halo in IFS data (the position of the AO cutoff frequency in the PSF is proportional to the wavelength) makes it easier to model close to the primary. In addition, the better AO performances in the NIR bands allow one to reach very good contrast and opens the path to detect new moons orbiting the primary vicinity, similarly to \Sthree.
Nonetheless, in this spectral range, the sky produces a non-negligible background signal, as seen in \subfig{fig:IP_overview}{e}, possibly hiding the faintest targets.

\textbf{ZIMPOL --} On the other side, without any internal heat production, these asteroids mainly reflect the sunlight and thus are slightly brighter in the visible band of ZIMPOL than in the NIR sensitivity range of IRDIS and the IFS. In addition, the sky background is negligible in this wavelength range \citep{Beuzit:19_SPHERE}. One can consequently hope to detect fainter (and thus smaller) moons orbiting asteroids at smaller separation with ZIMPOL. But at these smaller wavelengths, the AO system is less efficient and could leave a higher amount of scattered light around the primary. In addition the PSF of ZIMPOL is spatially more complicated than the PSF of the IFS since the AO cutoff frequency comes closer to the primary \citep[see \eg][]{Fetick:19_PSF_recons,Marchis:21_Kleopatra}. If the contrast is thus expected to be better far from the primary than in the IFS, the halo removal close to the asteroid remains a challenge where the IFS can provide complementary performances. This complementarity can also be useful to compare moons albedo with different instruments (visible \vs NIR) to provide insights on their surface chemical composition \citep{Reddy:12}.

\begin{figure}[ht!] 
        \centering
        
        \newcommand{\LineRatio}{0.56}
        
        \newcommand{\FigOne}{Med_projection_Elektra_2016-02-16_PIC.pdf}
        \newcommand{\FigTwo}{ZIMPOL_Elektra_2019-07-30-2_ZIMPOL.pdf}
        \newcommand{\FigThree}{ZIMPOL_Elektra_2019-08-06_ZIMPOL.pdf}
        \newcommand{\FigFour}{ZIMPOL_Color_bar.pdf}
        \newcommand{\subfigColor}{white}        
        
        \sbox1{\includegraphics{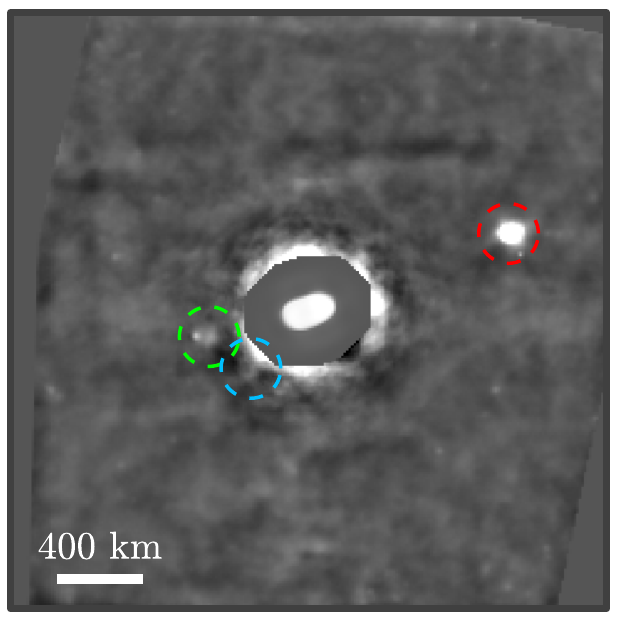}}               
        \sbox2{\includegraphics{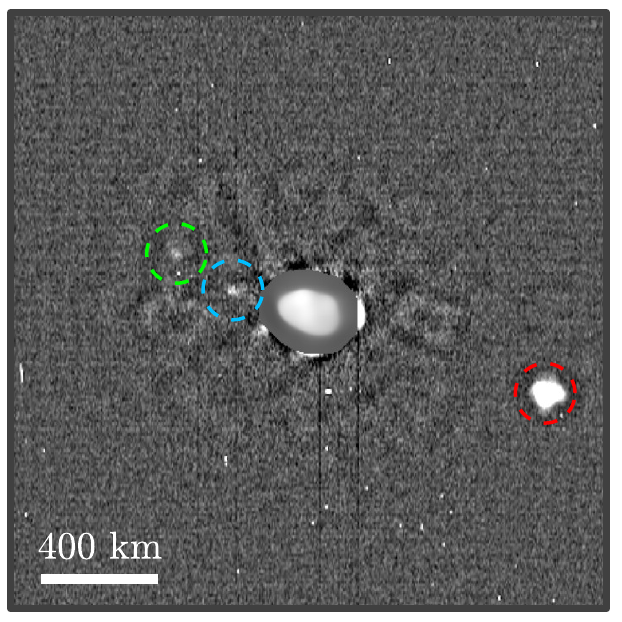}}               
        \sbox3{\includegraphics{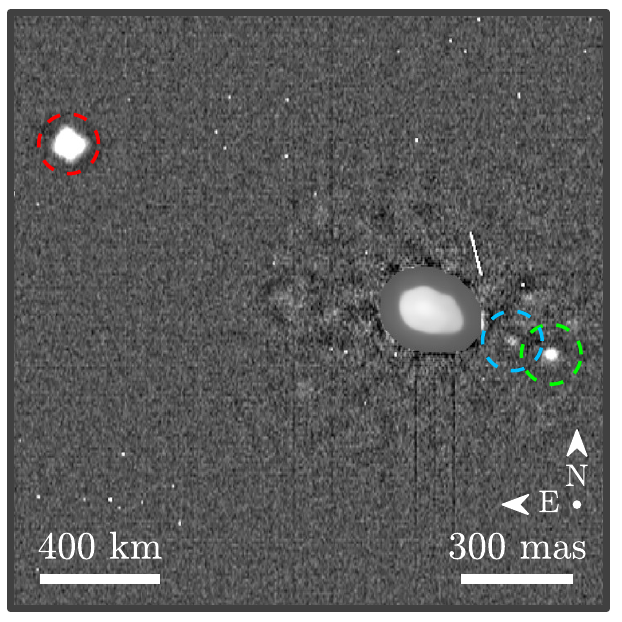}}     
        \sbox4{\includegraphics{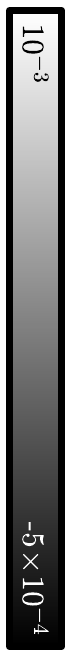}}              
        
        \newcommand{\ColumnWidth}[1]
                {\dimexpr \LineRatio \linewidth * \AspectRatio{#1} / (\AspectRatio{1} + \AspectRatio{4}) \relax
                }
        \newcommand{\ColumnGap}{\hspace {\dimexpr \linewidth /3 - \LineRatio\linewidth /3 }}

        \begin{tabular}{
                @{\ColumnGap}
                C{\ColumnWidth{1}}
                @{}
                C{\ColumnWidth{4}}
                @{\ColumnGap}
                }
                2016-02-16 / IFS &
                \\
                \subfigimg[width=\linewidth,pos=ul,font=\fontfig{\subfigColor}]{}{0.0}{\FigOne} &
                \subfigimg[width=\linewidth,pos=ul,font=\fontfig{\subfigColor}]{}{0.0}{\FigFour}
        \end{tabular}

        \vspace{-1pt}
        
        \renewcommand{\LineRatio}{1}
        \renewcommand{\ColumnWidth}[1]
                {\dimexpr \LineRatio \linewidth * \AspectRatio{#1} / (\AspectRatio{2} + \AspectRatio{3}) \relax
                }
        \renewcommand{\ColumnGap}{\hspace {\dimexpr \linewidth /3 - \LineRatio\linewidth /3 }}
        \begin{tabular}{
                @{\ColumnGap}
                C{\ColumnWidth{2}}
                @{\ColumnGap}
                C{\ColumnWidth{3}}
                @{\ColumnGap}
                }
                2019-07-30 / ZIMPOL &
                2019-08-06 / ZIMPOL
                \\
                \subfigimg[width=\linewidth,pos=ul,font=\fontfig{\subfigColor}]{}{0.0}{\FigTwo} &
                \subfigimg[width=\linewidth,pos=ul,font=\fontfig{\subfigColor}]{}{0.0}{\FigThree}
        \end{tabular}

        \caption{\label{fig:Extra_dates} Strong hints of detection in other datasets. The dashed circles indicate manually pinned positions on what could be a moon signal in the data. The central part of the field of view has been replaced by a dimmer deconvolved image of \Elektra to simultaneously visualise the primary and the moons. IFS: Median projection of 140 acquisitions. ZIMPOL: Single acquisition of each observation night.}
\end{figure}

Following the discovery of \Stwo, 140 additional SPHERE/IFS observations were performed on 16 Feb 2016 by Yang et al. (296.C-5038(A) -- `SPHERE Follow-up of the New Triple Asteroid (130) Elektra'). They were reduced with the PIC pipeline and processed as described in \refsec{sec:IP_overview}. Their temporal median projection is given in \fig{fig:Extra_dates}. The time-lapse of the data reduction and processing is given in {\textcolor{blue}{Visualisation 4}}. The conditions of observation were less favourable than in Dec 2014: the seeing was worse ($\sim\SI{1.1}{\arcsecond}$ versus  $\sim\SI{0.7}{\arcsecond}$ for 9 Dec 2014) and \Elektra was \SI{20}{\%} further along. We circle in \fig{fig:Extra_dates} what could be \Sthree without insuring that this is a detection.

Data were also gathered in Jul and Aug 2019 in the context of the systematic survey of main belt asteroids by \cite{Vernazza:21_survey_asteroid} with ZIMPOL (199.C-0074(E) -- `Asteroids as tracers of solar system formation: Probing the interior of primordial main belt asteroids'). As mentioned above, the halo removal algorithm presented in this Letter is not adapted to correctly model the halo shape in ZIMPOL data. We are currently working on a modified algorithm to account for the spatial features of the AO-corrected PSF. This is an on-going work and describing this method is beyond the scope of this Letter. This will be addressed in a dedicated forthcoming paper. We give very preliminary results in \fig{fig:Extra_dates} on two ZIMPOL acquisitions where the three moons are clearly visible.

We manually pinned strong hints of signal for the three moons in \fig{fig:Extra_dates}. These data are temporally too far from the 2014 acquisitions to be added to the orbital fit of \Sthree performed with Genoid. We present them to further support the discovery of \Sthree. The distance of the moon from \Elektra at the different epochs is consistent with the orbital parameters given in \tab{tab:Orb_fitting_S3} and discussed in \refsec{sec:Orb_fitting}.

In addition, we would like to mention here that for the 2016 epoch, the orbit planes of the moons are viewed edge-on. The angle between the orbits (adjusted on the 2014 observations) of \Sone and \Stwo and the orbit of \Sthree is given in \tab{tab:Spin_fitting} and is $38 \degr \pm 19 \degr$. The highlighted position in \fig{fig:Extra_dates} matches the geometric projection of such an inclined orbit for \Sthree. This suggests that this feature in the data could indeed be \Sthree. More careful data analyses are needed to confirm this suspicion, but this is beyond the scope of this Letter. Indeed, as mentioned above, the data quality at this epoch is poor.

\FloatBarrier

\section{Astrophotometric measurements}
\label{app:astrometry}

The tables in this appendix gather the astrophotometry fit for the three moons of \Elektra. In each table, $x$ and $y$ are the Cartesian positions of the moon relative to the photocentre of \Elektra and $\rho$ is the moon angular separation. Furetherore, $\delta m$ is the moon magnitude relative to the primary, and $D$ is the estimated moon diameter. The uncertainties on the positions are given on the top line of the tables for each date.

\longtab[1]{
    \begin{longtable}{|c|c|c|c|c|c|}
        \caption{Astrophotometry fit for \Sone{} -- \SoneFull.}
        \\ \hline
        9 Dec 2014 (Julian date) & $x \pm 4.3$ (\mas{}) & $y \pm 4.3$ (\mas{}) & $\rho \pm 4.3$ (\mas{}) & $\delta m \pm 0.20$  & $D$ (\SI{}{\kilo\meter}) \\ \hline
        2457000.56407 & 298.0 & 567.0 & 640 & 7.52 & 6.24 \\ \hline
        2457000.56446 & 297.0 & 567.0 & 640 & 7.56 & 6.12 \\ \hline
        2457000.56485 & 297.0 & 567.0 & 639 & 7.62 & 5.95 \\ \hline
        2457000.56525 & 296.0 & 566.0 & 639 & 7.60 & 6.01 \\ \hline
        2457000.56564 & 295.0 & 566.0 & 639 & 7.66 & 5.85 \\ \hline
        2457000.56609 & 295.0 & 566.0 & 639 & 7.65 & 5.87 \\ \hline
        2457000.56648 & 294.0 & 566.0 & 638 & 7.65 & 5.87 \\ \hline
        2457000.56688 & 294.0 & 566.0 & 638 & 7.66 & 5.85 \\ \hline
        2457000.56727 & 293.0 & 566.0 & 638 & 7.67 & 5.82 \\ \hline
        2457000.56766 & 293.0 & 566.0 & 637 & 7.66 & 5.85 \\ \hline
        2457000.56814 & 292.0 & 566.0 & 637 & 7.66 & 5.85 \\ \hline
        2457000.56853 & 292.0 & 566.0 & 637 & 7.67 & 5.82 \\ \hline
        2457000.56893 & 291.0 & 566.0 & 637 & 7.64 & 5.90 \\ \hline
        2457000.56932 & 291.0 & 566.0 & 636 & 7.66 & 5.85 \\ \hline
        2457000.56971 & 290.0 & 566.0 & 636 & 7.68 & 5.79 \\ \hline
        2457000.57016 & 290.0 & 566.0 & 636 & 7.65 & 5.87 \\ \hline
        2457000.57056 & 289.0 & 566.0 & 635 & 7.66 & 5.85 \\ \hline
        2457000.57095 & 288.0 & 566.0 & 635 & 7.68 & 5.79 \\ \hline
        2457000.57134 & 288.0 & 566.0 & 635 & 7.66 & 5.85 \\ \hline
        2457000.57174 & 287.0 & 566.0 & 635 & 7.65 & 5.87 \\ \hline
        2457000.57219 & 287.0 & 566.0 & 634 & 7.67 & 5.82 \\ \hline
        2457000.57258 & 286.0 & 566.0 & 634 & 7.63 & 5.93 \\ \hline
        2457000.57297 & 286.0 & 566.0 & 634 & 7.63 & 5.93 \\ \hline
        2457000.57337 & 285.0 & 565.0 & 633 & 7.63 & 5.93 \\ \hline
        2457000.57376 & 285.0 & 565.0 & 633 & 7.59 & 6.04 \\ \hline
        2457000.57422 & 284.0 & 565.0 & 633 & 7.57 & 6.09 \\ \hline
        2457000.57461 & 284.0 & 565.0 & 632 & 7.66 & 5.85 \\ \hline
        2457000.57500 & 283.0 & 565.0 & 632 & 7.61 & 5.98 \\ \hline
        2457000.57540 & 283.0 & 565.0 & 632 & 7.59 & 6.04 \\ \hline
        2457000.57579 & 282.0 & 565.0 & 632 & 7.60 & 6.01 \\ \hline
        2457000.57624 & 281.0 & 565.0 & 631 & 7.59 & 6.04 \\ \hline
        2457000.57663 & 281.0 & 565.0 & 631 & 7.60 & 6.01 \\ \hline
        2457000.57703 & 280.0 & 565.0 & 631 & 7.60 & 6.01 \\ \hline
        2457000.57742 & 280.0 & 565.0 & 630 & 7.59 & 6.04 \\ \hline
        2457000.57781 & 279.0 & 565.0 & 630 & 7.55 & 6.15 \\ \hline
        2457000.57826 & 279.0 & 565.0 & 630 & 7.59 & 6.04 \\ \hline
        2457000.57865 & 278.0 & 565.0 & 630 & 7.56 & 6.12 \\ \hline
        2457000.57904 & 278.0 & 565.0 & 629 & 7.58 & 6.07 \\ \hline
        2457000.57944 & 277.0 & 565.0 & 629 & 7.58 & 6.07 \\ \hline
        2457000.57983 & 277.0 & 565.0 & 629 & 7.56 & 6.12 \\ \hline
        2457000.58029 & 276.0 & 564.0 & 628 & 7.58 & 6.07 \\ \hline
        2457000.58068 & 276.0 & 564.0 & 628 & 7.53 & 6.21 \\ \hline
        2457000.58107 & 275.0 & 564.0 & 628 & 7.56 & 6.12 \\ \hline
        2457000.58147 & 275.0 & 564.0 & 628 & 7.55 & 6.15 \\ \hline
        2457000.58186 & 274.0 & 564.0 & 627 & 7.55 & 6.15 \\ \hline
        2457000.58232 & 274.0 & 564.0 & 627 & 7.58 & 6.07 \\ \hline
        2457000.58271 & 273.0 & 564.0 & 627 & 7.53 & 6.21 \\ \hline
        2457000.58310 & 272.0 & 564.0 & 626 & 7.54 & 6.18 \\ \hline
        2457000.58350 & 272.0 & 564.0 & 626 & 7.51 & 6.26 \\ \hline
        2457000.58389 & 271.0 & 564.0 & 626 & 7.54 & 6.18 \\ \hline
        \hline
        30 Dec 2014 (Julian date) & $x \pm 4.4$ (\mas{}) & $y \pm 4.4$ (\mas{}) & $\rho \pm 4.4$ (\mas{}) & $\delta m \pm 0.20$  & $D$ (\SI{}{\kilo\meter}) \\ \hline
        2457021.54396 & 473.0 & 472.0 & 668 & 7.52 & 6.24 \\ \hline
        2457021.54435 & 473.0 & 472.0 & 668 & 7.49 & 6.32 \\ \hline
        2457021.54475 & 472.0 & 472.0 & 667 & 7.50 & 6.29 \\ \hline
        2457021.54514 & 471.0 & 472.0 & 667 & 7.52 & 6.24 \\ \hline
        2457021.54553 & 470.0 & 472.0 & 666 & 7.53 & 6.21 \\ \hline
        2457021.54592 & 470.0 & 472.0 & 666 & 7.49 & 6.32 \\ \hline
        2457021.54632 & 469.0 & 472.0 & 665 & 7.49 & 6.32 \\ \hline
        2457021.54671 & 468.0 & 472.0 & 665 & 7.51 & 6.26 \\ \hline
        2457021.54710 & 467.0 & 472.0 & 664 & 7.53 & 6.21 \\ \hline
        2457021.54749 & 467.0 & 472.0 & 664 & 7.52 & 6.24 \\ \hline
        2457021.55448 & 466.0 & 472.0 & 663 & 7.52 & 6.24 \\ \hline
        2457021.55488 & 465.0 & 472.0 & 663 & 7.49 & 6.32 \\ \hline
        2457021.55527 & 464.0 & 472.0 & 662 & 7.54 & 6.18 \\ \hline
        2457021.55566 & 464.0 & 473.0 & 662 & 7.52 & 6.24 \\ \hline
        2457021.55605 & 463.0 & 473.0 & 661 & 7.51 & 6.26 \\ \hline
        2457021.55645 & 462.0 & 473.0 & 661 & 7.47 & 6.38 \\ \hline
        2457021.55684 & 461.0 & 473.0 & 660 & 7.48 & 6.35 \\ \hline
        2457021.55723 & 460.0 & 473.0 & 660 & 7.53 & 6.21 \\ \hline
        2457021.55762 & 460.0 & 473.0 & 660 & 7.44 & 6.47 \\ \hline
        2457021.55802 & 459.0 & 473.0 & 659 & 7.48 & 6.35 \\ \hline
        2457021.56081 & 458.0 & 473.0 & 659 & 7.51 & 6.26 \\ \hline
        2457021.56120 & 457.0 & 473.0 & 658 & 7.48 & 6.35 \\ \hline
        2457021.56159 & 457.0 & 473.0 & 658 & 7.49 & 6.32 \\ \hline
        2457021.56198 & 456.0 & 473.0 & 657 & 7.44 & 6.47 \\ \hline
        2457021.56238 & 455.0 & 473.0 & 657 & 7.51 & 6.26 \\ \hline
        2457021.56277 & 454.0 & 473.0 & 656 & 7.48 & 6.35 \\ \hline
        2457021.56316 & 454.0 & 473.0 & 656 & 7.50 & 6.29 \\ \hline
        2457021.56355 & 453.0 & 473.0 & 655 & 7.46 & 6.41 \\ \hline
        2457021.56395 & 452.0 & 473.0 & 655 & 7.48 & 6.35 \\ \hline
        2457021.56434 & 451.0 & 473.0 & 654 & 7.51 & 6.26 \\ \hline
        2457021.56709 & 451.0 & 474.0 & 654 & 7.54 & 6.18 \\ \hline
        2457021.56748 & 450.0 & 474.0 & 653 & 7.52 & 6.24 \\ \hline
        2457021.56788 & 449.0 & 474.0 & 653 & 7.49 & 6.32 \\ \hline
        2457021.56827 & 448.0 & 474.0 & 652 & 7.48 & 6.35 \\ \hline
        2457021.56866 & 448.0 & 474.0 & 652 & 7.49 & 6.32 \\ \hline
        2457021.56905 & 447.0 & 474.0 & 651 & 7.48 & 6.35 \\ \hline
        2457021.56945 & 446.0 & 474.0 & 651 & 7.51 & 6.26 \\ \hline
        2457021.56984 & 445.0 & 474.0 & 650 & 7.50 & 6.29 \\ \hline
        2457021.57023 & 445.0 & 474.0 & 650 & 7.48 & 6.35 \\ \hline
        2457021.57062 & 444.0 & 474.0 & 649 & 7.48 & 6.35 \\ \hline
        2457021.68937 & 309.0 & 478.0 & 569 & 7.71 & 5.71 \\ \hline
        2457021.68976 & 309.0 & 478.0 & 569 & 7.66 & 5.85 \\ \hline
        2457021.69016 & 308.0 & 478.0 & 568 & 7.58 & 6.07 \\ \hline
        2457021.69055 & 307.0 & 478.0 & 568 & 7.59 & 6.04 \\ \hline
        2457021.69094 & 307.0 & 478.0 & 568 & 7.53 & 6.21 \\ \hline
        2457021.69133 & 306.0 & 478.0 & 567 & 7.57 & 6.09 \\ \hline
        2457021.69173 & 305.0 & 477.0 & 567 & 7.57 & 6.09 \\ \hline
        2457021.69212 & 305.0 & 477.0 & 566 & 7.58 & 6.07 \\ \hline
        2457021.69251 & 304.0 & 477.0 & 566 & 7.57 & 6.09 \\ \hline
        2457021.69290 & 303.0 & 477.0 & 566 & 7.46 & 6.41 \\ \hline
        2457021.69563 & 303.0 & 477.0 & 565 & 7.56 & 6.12 \\ \hline
        2457021.69602 & 302.0 & 477.0 & 565 & 7.53 & 6.21 \\ \hline
        2457021.69642 & 301.0 & 477.0 & 564 & 7.52 & 6.24 \\ \hline
        2457021.69681 & 301.0 & 477.0 & 564 & 7.46 & 6.41 \\ \hline
        2457021.69720 & 300.0 & 477.0 & 564 & 7.52 & 6.24 \\ \hline
        2457021.69759 & 300.0 & 477.0 & 563 & 7.49 & 6.32 \\ \hline
        2457021.69799 & 299.0 & 477.0 & 563 & 7.44 & 6.47 \\ \hline
        2457021.69838 & 298.0 & 477.0 & 562 & 7.48 & 6.35 \\ \hline
        2457021.69877 & 298.0 & 477.0 & 562 & 7.46 & 6.41 \\ \hline
        2457021.69916 & 297.0 & 477.0 & 561 & 7.73 & 5.66 \\ \hline
        2457021.70188 & 296.0 & 476.0 & 561 & 7.66 & 5.85 \\ \hline
        2457021.70228 & 296.0 & 476.0 & 561 & 7.58 & 6.07 \\ \hline
        2457021.70267 & 295.0 & 476.0 & 560 & 7.60 & 6.01 \\ \hline
        2457021.70306 & 294.0 & 476.0 & 560 & 7.53 & 6.21 \\ \hline
        2457021.70345 & 294.0 & 476.0 & 559 & 7.57 & 6.09 \\ \hline
        2457021.70385 & 293.0 & 476.0 & 559 & 7.57 & 6.09 \\ \hline
        2457021.70424 & 292.0 & 476.0 & 559 & 7.56 & 6.12 \\ \hline
        2457021.70463 & 292.0 & 476.0 & 558 & 7.57 & 6.09 \\ \hline
        2457021.70502 & 291.0 & 476.0 & 558 & 7.49 & 6.32 \\ \hline
        2457021.70541 & 291.0 & 476.0 & 557 & 7.56 & 6.12 \\ \hline
        \hline
        31 Dec 2014 (Julian date) & $x \pm 4.4$ (\mas{}) & $y \pm 4.4$ (\mas{}) & $\rho \pm 4.4$ (\mas{}) & $\delta m \pm 0.20$  & $D$ (\SI{}{\kilo\meter}) \\ \hline
        2457022.65498 & -688.0 & 99.0 & 695 & 7.67 & 5.82 \\ \hline
        2457022.65537 & -688.0 & 99.0 & 695 & 7.72 & 5.69 \\ \hline
        2457022.65576 & -688.0 & 98.0 & 695 & 7.67 & 5.82 \\ \hline
        2457022.65615 & -689.0 & 98.0 & 696 & 7.71 & 5.71 \\ \hline
        2457022.65655 & -689.0 & 98.0 & 696 & 7.71 & 5.71 \\ \hline
        2457022.65694 & -689.0 & 97.0 & 696 & 7.70 & 5.74 \\ \hline
        2457022.65733 & -689.0 & 97.0 & 696 & 7.74 & 5.63 \\ \hline
        2457022.65772 & -690.0 & 96.0 & 696 & 7.74 & 5.63 \\ \hline
        2457022.65812 & -690.0 & 96.0 & 697 & 7.65 & 5.87 \\ \hline
        2457022.65851 & -690.0 & 95.0 & 697 & 7.73 & 5.66 \\ \hline
        2457022.66125 & -691.0 & 95.0 & 697 & 7.66 & 5.85 \\ \hline
        2457022.66164 & -691.0 & 94.0 & 697 & 7.58 & 6.07 \\ \hline
        2457022.66204 & -691.0 & 94.0 & 698 & 7.60 & 6.01 \\ \hline
        2457022.66243 & -692.0 & 94.0 & 698 & 7.53 & 6.21 \\ \hline
        2457022.66282 & -692.0 & 93.0 & 698 & 7.57 & 6.09 \\ \hline
        2457022.66321 & -692.0 & 93.0 & 698 & 7.57 & 6.09 \\ \hline
        2457022.66361 & -693.0 & 92.0 & 699 & 7.58 & 6.07 \\ \hline
        2457022.66400 & -693.0 & 92.0 & 699 & 7.57 & 6.09 \\ \hline
        2457022.66439 & -693.0 & 91.0 & 699 & 7.49 & 6.32 \\ \hline
        2457022.66478 & -693.0 & 91.0 & 699 & 7.56 & 6.12 \\ \hline
        2457022.66750 & -694.0 & 90.0 & 700 & 7.53 & 6.21 \\ \hline
        2457022.66789 & -694.0 & 90.0 & 700 & 7.52 & 6.24 \\ \hline
        2457022.66828 & -694.0 & 89.0 & 700 & 7.46 & 6.41 \\ \hline
        2457022.66867 & -695.0 & 89.0 & 700 & 7.52 & 6.24 \\ \hline
        2457022.66907 & -695.0 & 88.0 & 701 & 7.49 & 6.32 \\ \hline
        2457022.66946 & -695.0 & 88.0 & 701 & 7.42 & 6.53 \\ \hline
        2457022.66985 & -696.0 & 88.0 & 701 & 7.48 & 6.35 \\ \hline
        2457022.67024 & -696.0 & 87.0 & 701 & 7.46 & 6.41 \\ \hline
        2457022.67064 & -696.0 & 87.0 & 702 & 7.48 & 6.35 \\ \hline
        2457022.67103 & -696.0 & 86.0 & 702 & 7.48 & 6.35 \\ \hline
    \end{longtable}

    \begin{longtable}{|c|c|c|c|c|c|}
        \caption{Astrophotometry fit for \Stwo{} -- \StwoFull.}
        \\ \hline
        9 Dec 2014 (Julian date) & $x \pm 4.3$ (\mas{}) & $y \pm 4.3$ (\mas{}) & $\rho \pm 4.3$ (\mas{}) & $\delta m \pm 0.30$  & $D$ (\SI{}{\kilo\meter}) \\ \hline
        2457000.56407 & 356.3 & 8.7 & 356 & 9.76 & 2.22 \\ \hline
        2457000.56446 & 356.5 & 9.2 & 357 & 9.82 & 2.16 \\ \hline
        2457000.56485 & 356.6 & 9.8 & 357 & 9.98 & 2.01 \\ \hline
        2457000.56525 & 356.8 & 10.3 & 357 & 9.80 & 2.18 \\ \hline
        2457000.56564 & 357.0 & 10.9 & 357 & 10.12 & 1.88 \\ \hline
        2457000.56609 & 357.2 & 11.4 & 357 & 10.00 & 1.99 \\ \hline
        2457000.56648 & 357.4 & 12.0 & 358 & 10.03 & 1.96 \\ \hline
        2457000.56688 & 357.6 & 12.5 & 358 & 10.32 & 1.72 \\ \hline
        2457000.56727 & 357.8 & 13.1 & 358 & 10.07 & 1.93 \\ \hline
        2457000.56766 & 358.0 & 13.6 & 358 & 9.97 & 2.02 \\ \hline
        2457000.56814 & 358.2 & 14.2 & 359 & 10.06 & 1.94 \\ \hline
        2457000.56853 & 358.4 & 14.8 & 359 & 10.02 & 1.97 \\ \hline
        2457000.56893 & 358.6 & 15.3 & 359 & 9.96 & 2.03 \\ \hline
        2457000.56932 & 358.7 & 15.9 & 359 & 9.86 & 2.12 \\ \hline
        2457000.56971 & 358.9 & 16.4 & 359 & 9.93 & 2.06 \\ \hline
        2457000.57016 & 359.1 & 17.0 & 360 & 9.97 & 2.02 \\ \hline
        2457000.57056 & 359.3 & 17.5 & 360 & 10.12 & 1.88 \\ \hline
        2457000.57095 & 359.5 & 18.1 & 360 & 10.06 & 1.94 \\ \hline
        2457000.57134 & 359.7 & 18.6 & 360 & 9.99 & 2.00 \\ \hline
        2457000.57174 & 359.8 & 19.2 & 360 & 10.06 & 1.94 \\ \hline
        2457000.57219 & 360.0 & 19.8 & 361 & 9.96 & 2.03 \\ \hline
        2457000.57258 & 360.2 & 20.3 & 361 & 10.24 & 1.78 \\ \hline
        2457000.57297 & 360.4 & 20.9 & 361 & 10.03 & 1.96 \\ \hline
        2457000.57337 & 360.5 & 21.4 & 361 & 10.10 & 1.90 \\ \hline
        2457000.57376 & 360.7 & 22.0 & 361 & 10.10 & 1.90 \\ \hline
        2457000.57422 & 360.9 & 22.6 & 362 & 10.03 & 1.96 \\ \hline
        2457000.57461 & 361.1 & 23.1 & 362 & 10.21 & 1.81 \\ \hline
        2457000.57500 & 361.2 & 23.7 & 362 & 9.98 & 2.01 \\ \hline
        2457000.57540 & 361.4 & 24.2 & 362 & 9.93 & 2.06 \\ \hline
        2457000.57579 & 361.6 & 24.8 & 362 & 10.05 & 1.94 \\ \hline
        2457000.57624 & 361.8 & 25.4 & 363 & 10.17 & 1.84 \\ \hline
        2457000.57663 & 361.9 & 25.9 & 363 & 10.01 & 1.98 \\ \hline
        2457000.57703 & 362.1 & 26.5 & 363 & 9.95 & 2.04 \\ \hline
        2457000.57742 & 362.3 & 27.1 & 363 & 10.23 & 1.79 \\ \hline
        2457000.57781 & 362.4 & 27.6 & 364 & 10.07 & 1.93 \\ \hline
        2457000.57826 & 362.6 & 28.2 & 364 & 10.13 & 1.87 \\ \hline
        2457000.57865 & 362.8 & 28.8 & 364 & 10.10 & 1.90 \\ \hline
        2457000.57904 & 362.9 & 29.3 & 364 & 10.12 & 1.88 \\ \hline
        2457000.57944 & 363.1 & 29.9 & 364 & 9.97 & 2.02 \\ \hline
        2457000.57983 & 363.3 & 30.5 & 365 & 10.00 & 1.99 \\ \hline
        2457000.58029 & 363.4 & 31.1 & 365 & 9.88 & 2.10 \\ \hline
        2457000.58068 & 363.6 & 31.6 & 365 & 10.00 & 1.99 \\ \hline
        2457000.58107 & 363.8 & 32.2 & 365 & 9.96 & 2.03 \\ \hline
        2457000.58147 & 363.9 & 32.8 & 365 & 9.77 & 2.21 \\ \hline
        2457000.58186 & 364.1 & 33.3 & 366 & 10.01 & 1.98 \\ \hline
        2457000.58232 & 364.2 & 33.9 & 366 & 10.02 & 1.97 \\ \hline
        2457000.58271 & 364.4 & 34.5 & 366 & 10.17 & 1.84 \\ \hline
        2457000.58310 & 364.5 & 35.1 & 366 & 10.03 & 1.96 \\ \hline
        2457000.58350 & 364.7 & 35.6 & 366 & 10.11 & 1.89 \\ \hline
        2457000.58389 & 364.8 & 36.2 & 367 & 9.94 & 2.05 \\ \hline
        \hline
        30 Dec 2014 (Julian date) & $x \pm 4.5$ (\mas{}) & $y \pm 4.5$ (\mas{}) & $\rho \pm 4.5$ (\mas{}) & $\delta m \pm 0.40$  & $D$ (\SI{}{\kilo\meter}) \\ \hline
        2457021.54396 & -317.9 & -112.2 & 337 & 9.82 & 2.16 \\ \hline
        2457021.54435 & -317.5 & -113.1 & 337 & 10.31 & 1.73 \\ \hline
        2457021.54475 & -317.2 & -113.9 & 337 & 10.24 & 1.78 \\ \hline
        2457021.54514 & -316.8 & -114.8 & 337 & 10.06 & 1.94 \\ \hline
        2457021.54553 & -316.4 & -115.6 & 337 & 10.06 & 1.94 \\ \hline
        2457021.54592 & -316.1 & -116.5 & 337 & 10.17 & 1.84 \\ \hline
        2457021.54632 & -315.7 & -117.3 & 337 & 10.17 & 1.84 \\ \hline
        2457021.54671 & -315.3 & -118.1 & 337 & 10.09 & 1.91 \\ \hline
        2457021.54710 & -315.0 & -119.0 & 337 & 10.06 & 1.94 \\ \hline
        2457021.54749 & -314.6 & -119.8 & 337 & 10.15 & 1.86 \\ \hline
        2457021.55448 & -314.2 & -120.6 & 337 & 10.36 & 1.69 \\ \hline
        2457021.55488 & -313.8 & -121.5 & 337 & 10.07 & 1.93 \\ \hline
        2457021.55527 & -313.4 & -122.3 & 336 & 10.38 & 1.67 \\ \hline
        2457021.55566 & -313.1 & -123.1 & 336 & 9.99 & 2.00 \\ \hline
        2457021.55605 & -312.7 & -124.0 & 336 & 10.07 & 1.93 \\ \hline
        2457021.55645 & -312.3 & -124.8 & 336 & 10.14 & 1.87 \\ \hline
        2457021.55684 & -311.9 & -125.6 & 336 & 9.94 & 2.05 \\ \hline
        2457021.55723 & -311.5 & -126.4 & 336 & 10.35 & 1.69 \\ \hline
        2457021.55762 & -311.1 & -127.3 & 336 & 10.24 & 1.78 \\ \hline
        2457021.55802 & -310.7 & -128.1 & 336 & 10.19 & 1.82 \\ \hline
        2457021.56081 & -310.3 & -128.9 & 336 & 10.14 & 1.87 \\ \hline
        2457021.56120 & -309.9 & -129.7 & 336 & 10.38 & 1.67 \\ \hline
        2457021.56159 & -309.5 & -130.5 & 336 & 10.04 & 1.95 \\ \hline
        2457021.56198 & -309.1 & -131.4 & 336 & 10.13 & 1.87 \\ \hline
        2457021.56238 & -308.7 & -132.2 & 336 & 10.10 & 1.90 \\ \hline
        2457021.56277 & -308.3 & -133.0 & 336 & 10.31 & 1.73 \\ \hline
        2457021.56316 & -307.9 & -133.8 & 336 & 10.04 & 1.95 \\ \hline
        2457021.56355 & -307.4 & -134.6 & 336 & 10.13 & 1.87 \\ \hline
        2457021.56395 & -307.0 & -135.4 & 336 & 9.89 & 2.09 \\ \hline
        2457021.56434 & -306.6 & -136.2 & 336 & 9.93 & 2.06 \\ \hline
        2457021.56709 & -306.2 & -137.1 & 335 & 9.86 & 2.12 \\ \hline
        2457021.56748 & -305.8 & -137.9 & 335 & 10.03 & 1.96 \\ \hline
        2457021.56788 & -305.3 & -138.7 & 335 & 9.72 & 2.26 \\ \hline
        2457021.56827 & -304.9 & -139.5 & 335 & 9.65 & 2.34 \\ \hline
        2457021.56866 & -304.5 & -140.3 & 335 & 9.64 & 2.35 \\ \hline
        2457021.56905 & -304.1 & -141.1 & 335 & 9.62 & 2.37 \\ \hline
        2457021.56945 & -303.6 & -141.9 & 335 & 9.84 & 2.14 \\ \hline
        2457021.56984 & -303.2 & -142.7 & 335 & 9.77 & 2.21 \\ \hline
        2457021.57023 & -302.7 & -143.5 & 335 & 9.77 & 2.21 \\ \hline
        2457021.57062 & -302.3 & -144.3 & 335 & 9.79 & 2.19 \\ \hline
        \hline
        31 Dec 2014 (Julian date) & $x \pm 5.2$ (\mas{}) & $y \pm 5.2$ (\mas{}) & $\rho \pm 5.2$ (\mas{}) & $\delta m \pm 0.40$  & $D$ (\SI{}{\kilo\meter}) \\ \hline
        2457022.65498 & -314.1 & -37.8 & 316 & 10.44 & 1.62 \\ \hline
        2457022.65537 & -314.2 & -38.3 & 317 & 10.43 & 1.63 \\ \hline
        2457022.65576 & -314.4 & -38.7 & 317 & 10.39 & 1.66 \\ \hline
        2457022.65615 & -314.6 & -39.2 & 317 & 10.33 & 1.71 \\ \hline
        2457022.65655 & -314.8 & -39.7 & 317 & 10.19 & 1.82 \\ \hline
        2457022.65694 & -315.0 & -40.2 & 318 & 10.00 & 1.99 \\ \hline
        2457022.65733 & -315.2 & -40.7 & 318 & 10.26 & 1.77 \\ \hline
        2457022.65772 & -315.3 & -41.2 & 318 & 10.11 & 1.89 \\ \hline
        2457022.65812 & -315.5 & -41.6 & 318 & 10.64 & 1.48 \\ \hline
        2457022.65851 & -315.7 & -42.1 & 318 & 10.13 & 1.87 \\ \hline
        2457022.66125 & -315.9 & -42.6 & 319 & 9.86 & 2.12 \\ \hline
        2457022.66164 & -316.0 & -43.1 & 319 & 9.95 & 2.04 \\ \hline
        2457022.66204 & -316.2 & -43.6 & 319 & 9.96 & 2.03 \\ \hline
        2457022.66243 & -316.4 & -44.1 & 319 & 9.85 & 2.13 \\ \hline
        2457022.66282 & -316.6 & -44.6 & 320 & 10.09 & 1.91 \\ \hline
        2457022.66321 & -316.7 & -45.1 & 320 & 10.05 & 1.94 \\ \hline
        2457022.66361 & -316.9 & -45.6 & 320 & 10.02 & 1.97 \\ \hline
        2457022.66400 & -317.1 & -46.0 & 320 & 9.93 & 2.06 \\ \hline
        2457022.66439 & -317.3 & -46.5 & 321 & 9.95 & 2.04 \\ \hline
        2457022.66478 & -317.4 & -47.0 & 321 & 10.02 & 1.97 \\ \hline
        2457022.66750 & -317.6 & -47.5 & 321 & 10.17 & 1.84 \\ \hline
        2457022.66789 & -317.8 & -48.0 & 321 & 10.47 & 1.60 \\ \hline
        2457022.66828 & -317.9 & -48.5 & 322 & 9.95 & 2.04 \\ \hline
        2457022.66867 & -318.1 & -49.0 & 322 & 10.49 & 1.59 \\ \hline
        2457022.66907 & -318.3 & -49.5 & 322 & 9.94 & 2.05 \\ \hline
        2457022.66946 & -318.4 & -50.0 & 322 & 10.03 & 1.96 \\ \hline
        2457022.66985 & -318.6 & -50.5 & 323 & 10.11 & 1.89 \\ \hline
        2457022.67024 & -318.8 & -51.0 & 323 & 10.05 & 1.94 \\ \hline
        2457022.67064 & -318.9 & -51.5 & 323 & 10.27 & 1.76 \\ \hline
        2457022.67103 & -319.1 & -52.0 & 323 & 9.71 & 2.27 \\ \hline
    \end{longtable}

    \begin{longtable}{|c|c|c|c|c|c|}
        \caption{Astrophotometry fit for \Sthree{} -- \SthreeFull.}
        \\ \hline
        9 Dec 2014 (Julian date) & $x \pm 6.0$ (\mas{}) & $y \pm 6.0$ (\mas{}) & $\rho \pm 6.0$ (\mas{}) & $\delta m \pm 0.50$  & $D$ (\SI{}{\kilo\meter}) \\ \hline
        2457000.56407 & 55.8 & -253.4 & 259 & 10.26 & 1.77 \\ \hline
        2457000.56446 & 56.4 & -253.2 & 259 & 10.45 & 1.62 \\ \hline
        2457000.56485 & 57.0 & -253.0 & 259 & 10.69 & 1.45 \\ \hline
        2457000.56525 & 57.6 & -252.9 & 259 & 10.39 & 1.66 \\ \hline
        2457000.56564 & 58.2 & -252.7 & 259 & 10.66 & 1.47 \\ \hline
        2457000.56609 & 58.8 & -252.5 & 259 & 10.56 & 1.54 \\ \hline
        2457000.56648 & 59.4 & -252.3 & 259 & 10.36 & 1.69 \\ \hline
        2457000.56688 & 60.0 & -252.1 & 259 & 10.43 & 1.63 \\ \hline
        2457000.56727 & 60.7 & -251.9 & 259 & 10.60 & 1.51 \\ \hline
        2457000.56766 & 61.3 & -251.7 & 259 & 10.41 & 1.65 \\ \hline
        2457000.56814 & 61.8 & -251.5 & 259 & 10.54 & 1.55 \\ \hline
        2457000.56853 & 62.4 & -251.3 & 259 & 10.65 & 1.48 \\ \hline
        2457000.56893 & 63.0 & -251.1 & 259 & 10.39 & 1.66 \\ \hline
        2457000.56932 & 63.6 & -250.9 & 259 & 10.39 & 1.66 \\ \hline
        2457000.56971 & 64.2 & -250.7 & 259 & 9.80 & 2.18 \\ \hline
        2457000.57016 & 64.8 & -250.5 & 259 & 10.64 & 1.48 \\ \hline
        2457000.57056 & 65.4 & -250.3 & 259 & 10.54 & 1.55 \\ \hline
        2457000.57095 & 66.0 & -250.1 & 259 & 10.76 & 1.40 \\ \hline
        2457000.57134 & 66.6 & -249.9 & 259 & 10.64 & 1.48 \\ \hline
        2457000.57174 & 67.2 & -249.7 & 259 & 10.75 & 1.41 \\ \hline
        2457000.57219 & 67.8 & -249.5 & 259 & 10.41 & 1.65 \\ \hline
        2457000.57258 & 68.4 & -249.2 & 258 & 10.56 & 1.54 \\ \hline
        2457000.57297 & 69.0 & -249.0 & 258 & 10.27 & 1.76 \\ \hline
        2457000.57337 & 69.6 & -248.8 & 258 & 10.52 & 1.57 \\ \hline
        2457000.57376 & 70.2 & -248.6 & 258 & 10.61 & 1.50 \\ \hline
        2457000.57422 & 70.8 & -248.4 & 258 & 10.43 & 1.63 \\ \hline
        2457000.57461 & 71.4 & -248.2 & 258 & 10.48 & 1.60 \\ \hline
        2457000.57500 & 72.0 & -247.9 & 258 & 10.40 & 1.66 \\ \hline
        2457000.57540 & 72.5 & -247.7 & 258 & 10.37 & 1.68 \\ \hline
        2457000.57579 & 73.1 & -247.5 & 258 & 10.46 & 1.61 \\ \hline
        2457000.57624 & 73.7 & -247.3 & 258 & 10.51 & 1.57 \\ \hline
        2457000.57663 & 74.3 & -247.0 & 258 & 10.28 & 1.75 \\ \hline
        2457000.57703 & 74.9 & -246.8 & 258 & 10.56 & 1.54 \\ \hline
        2457000.57742 & 75.5 & -246.6 & 258 & 10.47 & 1.60 \\ \hline
        2457000.57781 & 76.1 & -246.3 & 258 & 10.33 & 1.71 \\ \hline
        2457000.57826 & 76.6 & -246.1 & 258 & 10.64 & 1.48 \\ \hline
        2457000.57865 & 77.2 & -245.9 & 258 & 10.39 & 1.66 \\ \hline
        2457000.57904 & 77.8 & -245.6 & 258 & 10.67 & 1.46 \\ \hline
        2457000.57944 & 78.4 & -245.4 & 258 & 10.54 & 1.55 \\ \hline
        2457000.57983 & 79.0 & -245.2 & 258 & 10.41 & 1.65 \\ \hline
        2457000.58029 & 79.5 & -244.9 & 258 & 10.20 & 1.81 \\ \hline
        2457000.58068 & 80.1 & -244.7 & 257 & 10.49 & 1.59 \\ \hline
        2457000.58107 & 80.7 & -244.5 & 257 & 10.76 & 1.40 \\ \hline
        2457000.58147 & 81.3 & -244.2 & 257 & 11.07 & 1.22 \\ \hline
        2457000.58186 & 81.9 & -244.0 & 257 & 10.59 & 1.52 \\ \hline
        2457000.58232 & 82.4 & -243.7 & 257 & 10.61 & 1.50 \\ \hline
        2457000.58271 & 83.0 & -243.5 & 257 & 10.48 & 1.60 \\ \hline
        2457000.58310 & 83.6 & -243.2 & 257 & 10.86 & 1.34 \\ \hline
        2457000.58350 & 84.2 & -243.0 & 257 & 10.59 & 1.52 \\ \hline
        2457000.58389 & 84.7 & -242.7 & 257 & 10.68 & 1.45 \\ \hline
        \hline
        30 Dec 2014 (Julian date) & $x \pm 7.1$ (\mas{}) & $y \pm 7.1$ (\mas{}) & $\rho \pm 7.1$ (\mas{}) & $\delta m \pm 0.50$  & $D$ (\SI{}{\kilo\meter}) \\ \hline
        2457021.54396 & -70.7 & -233.3 & 244 & 10.51 & 1.57 \\ \hline
        2457021.54435 & -69.4 & -232.9 & 243 & 10.60 & 1.51 \\ \hline
        2457021.54475 & -68.1 & -232.5 & 242 & 9.89 & 2.09 \\ \hline
        2457021.54514 & -66.8 & -232.1 & 242 & 11.25 & 1.12 \\ \hline
        2457021.54553 & -65.5 & -231.7 & 241 & 10.47 & 1.60 \\ \hline
        2457021.54592 & -64.2 & -231.3 & 240 & 10.39 & 1.66 \\ \hline
        2457021.54632 & -62.9 & -230.8 & 239 & 10.89 & 1.32 \\ \hline
        2457021.54671 & -61.7 & -230.4 & 239 & 10.66 & 1.47 \\ \hline
        2457021.54710 & -60.4 & -230.0 & 238 & 10.44 & 1.62 \\ \hline
        2457021.54749 & -59.1 & -229.5 & 237 & 10.50 & 1.58 \\ \hline
        2457021.55448 & -57.9 & -229.1 & 236 & 9.95 & 2.04 \\ \hline
        2457021.55488 & -56.6 & -228.6 & 236 & 10.20 & 1.81 \\ \hline
        2457021.55527 & -55.4 & -228.2 & 235 & 10.17 & 1.84 \\ \hline
        2457021.55566 & -54.1 & -227.7 & 234 & 10.30 & 1.73 \\ \hline
        2457021.55605 & -52.9 & -227.2 & 233 & 10.32 & 1.72 \\ \hline
        2457021.55645 & -51.6 & -226.7 & 233 & 10.26 & 1.77 \\ \hline
        2457021.55684 & -50.4 & -226.2 & 232 & 10.35 & 1.69 \\ \hline
        2457021.55723 & -49.2 & -225.7 & 231 & 10.43 & 1.63 \\ \hline
        2457021.55762 & -48.0 & -225.2 & 230 & 10.23 & 1.79 \\ \hline
        2457021.55802 & -46.8 & -224.7 & 230 & 9.98 & 2.01 \\ \hline
        2457021.56081 & -45.6 & -224.2 & 229 & 10.26 & 1.77 \\ \hline
        2457021.56120 & -44.4 & -223.7 & 228 & 9.79 & 2.19 \\ \hline
        2457021.56159 & -43.2 & -223.2 & 227 & 10.08 & 1.92 \\ \hline
        2457021.56198 & -42.0 & -222.6 & 227 & 9.99 & 2.00 \\ \hline
        2457021.56238 & -40.8 & -222.1 & 226 & 10.29 & 1.74 \\ \hline
        2457021.56277 & -39.6 & -221.5 & 225 & 9.71 & 2.27 \\ \hline
        2457021.56316 & -38.5 & -221.0 & 224 & 9.96 & 2.03 \\ \hline
        2457021.56355 & -37.3 & -220.4 & 224 & 10.23 & 1.79 \\ \hline
        2457021.56395 & -36.1 & -219.9 & 223 & 10.32 & 1.72 \\ \hline
        2457021.56434 & -35.0 & -219.3 & 222 & 9.99 & 2.00 \\ \hline
        2457021.56709 & -33.8 & -218.7 & 221 & 10.24 & 1.78 \\ \hline
        2457021.56748 & -32.7 & -218.1 & 221 & 9.98 & 2.01 \\ \hline
        2457021.56788 & -31.6 & -217.5 & 220 & 9.41 & 2.61 \\ \hline
        2457021.56827 & -30.4 & -216.9 & 219 & 9.56 & 2.44 \\ \hline
        2457021.56866 & -29.3 & -216.3 & 218 & 9.67 & 2.32 \\ \hline
        2457021.56905 & -28.2 & -215.7 & 218 & 10.30 & 1.73 \\ \hline
        2457021.56945 & -27.1 & -215.1 & 217 & 10.02 & 1.97 \\ \hline
        2457021.56984 & -26.0 & -214.5 & 216 & 9.98 & 2.01 \\ \hline
        2457021.57023 & -24.9 & -213.9 & 215 & 9.94 & 2.05 \\ \hline
        2457021.57062 & -23.8 & -213.3 & 215 & 10.02 & 1.97 \\ \hline
        \hline
        31 Dec 2014 (Julian date) & $x \pm 6.2$ (\mas{}) & $y \pm 6.2$ (\mas{}) & $\rho \pm 6.2$ (\mas{}) & $\delta m \pm 0.50$  & $D$ (\SI{}{\kilo\meter}) \\ \hline
        2457022.65498 & -312.8 & 14.0 & 313 & 11.10 & 1.20 \\ \hline
        2457022.65537 & -313.2 & 13.8 & 314 & 10.67 & 1.46 \\ \hline
        2457022.65576 & -313.6 & 13.6 & 314 & 10.21 & 1.81 \\ \hline
        2457022.65615 & -314.0 & 13.4 & 314 & 10.74 & 1.42 \\ \hline
        2457022.65655 & -314.4 & 13.2 & 315 & 10.05 & 1.94 \\ \hline
        2457022.65694 & -314.8 & 13.0 & 315 & 10.37 & 1.68 \\ \hline
        2457022.65733 & -315.2 & 12.7 & 316 & 10.53 & 1.56 \\ \hline
        2457022.65772 & -315.7 & 12.5 & 316 & 10.15 & 1.86 \\ \hline
        2457022.65812 & -316.1 & 12.3 & 316 & 10.85 & 1.35 \\ \hline
        2457022.65851 & -316.5 & 12.1 & 317 & 10.83 & 1.36 \\ \hline
        2457022.66125 & -316.9 & 11.9 & 317 & 10.36 & 1.69 \\ \hline
        2457022.66164 & -317.3 & 11.7 & 318 & 10.24 & 1.78 \\ \hline
        2457022.66204 & -317.7 & 11.4 & 318 & 10.36 & 1.69 \\ \hline
        2457022.66243 & -318.1 & 11.2 & 318 & 10.34 & 1.70 \\ \hline
        2457022.66282 & -318.5 & 11.0 & 319 & 10.15 & 1.86 \\ \hline
        2457022.66321 & -318.9 & 10.8 & 319 & 10.13 & 1.87 \\ \hline
        2457022.66361 & -319.3 & 10.6 & 320 & 10.28 & 1.75 \\ \hline
        2457022.66400 & -319.7 & 10.3 & 320 & 10.17 & 1.84 \\ \hline
        2457022.66439 & -320.1 & 10.1 & 320 & 10.38 & 1.67 \\ \hline
        2457022.66478 & -320.5 & 9.9 & 321 & 10.45 & 1.62 \\ \hline
        2457022.66750 & -321.0 & 9.7 & 321 & 10.62 & 1.50 \\ \hline
        2457022.66789 & -321.4 & 9.4 & 322 & 10.30 & 1.73 \\ \hline
        2457022.66828 & -321.8 & 9.2 & 322 & 10.20 & 1.81 \\ \hline
        2457022.66867 & -322.2 & 9.0 & 322 & 10.28 & 1.75 \\ \hline
        2457022.66907 & -322.6 & 8.8 & 323 & 10.22 & 1.80 \\ \hline
        2457022.66946 & -323.0 & 8.5 & 323 & 10.31 & 1.73 \\ \hline
        2457022.66985 & -323.4 & 8.3 & 324 & 10.98 & 1.27 \\ \hline
        2457022.67024 & -323.8 & 8.1 & 324 & 11.55 & 0.97 \\ \hline
        2457022.67064 & -324.2 & 7.9 & 324 & 10.56 & 1.54 \\ \hline
        2457022.67103 & -324.6 & 7.6 & 325 & 10.56 & 1.54 \\ \hline
    \end{longtable}
}

\end{appendix}

\end{document}